\documentclass[letterpaper]{article}
\usepackage{color}
\usepackage{graphicx}

\begin{document}
\title{Single charge-current in a normal mesoscopic region attached to
       superconductor leads via a coupled Poisson Non-equilibrium Green Function formalism}

\author{David\ Verrilli${}^{1}$, F.\ P.\ Marin${}^{1}$,  Rafael\ Rangel${}^{2}$\\
         ${}^{1}$ Laboratorio de F\'{\i}sica Te\'{o}rica de S\'{o}lidos (LFTS).\\
         Centro de F\'{\i}sica Te\'{o}rica y Computacional (CEFITEC).\\
         Facultad de Ciencias. Universidad Central de Venezuela.\\
         A.P. 47586. Caracas 1041-A. Venezuela, \\
         ${}^{2}$ Departamento de F\'{\i}sica.\\
         Universidad Sim\'{\o}n  Bol\'{\i}var.\\
         A.P. 89000. Caracas 1080-A. Venezuela}

\maketitle

We study the ${\rm I}-V$ characteristic of a mesoscopic systems  or quantum dot (QD) attached to a pair of superconducting leads.
Interaction effects in the QD are considered through  the charging energy of the QD, i.e., the treatment of current transport  under  a
voltage bias is performed within a coupled Poisson Non-equilibrium Green Function (PNEGF) formalism. We derive the expression for the
current in full generality, but  consider only the regime where transport occurs only via a single particle current. We  show for this case
and for  various charging energies values $U_{\rm 0}$ and associated capacitances of the QD , the effect  on the ${\rm I}-V$ characteristic.
Also the influence of the coupling constants on the  ${\rm I}-V$ characteristic is investigated. Our approach puts forward a novel
interpretation of experiments in the strong  Coulomb regime.

\section{Introduction}\label{intro}
The overall shape of the ${\rm I}-V$ characteristic of a variety of systems (metals, semiconductors, molecular conductors) in the nanometer
scale sandwiched between metallic or superconductors leads has been recently a matter of study (see
\cite{KouwenhofenReport,DeFranceshiReview} and references there in). In these systems, the energy level discreteness is quite important
since level spacing is comparable with  other energy scales \cite{MarcKastner,SDattaBook2005}. Indeed,
the coupling with the bath modifies drastically the properties of an otherwise uncoupled nanometer system in a sharp
contrast with similar non-equilibrium macroscopic systems
\cite{RTidecks,MTinkham,NKopnin,Keldysh2003,RvaanLwewen,FPMarin,PekolaKlapwijk,SerbynSkvortsov,SnymanNazarov}. They constitutes hybrid
systems. Theoretical studies \cite{FClaro,MKrawiec,KKang,MRoderoI,MRoderoII,JKoenig,Giazotto} as well as experimental measurements have been
done by many research groups \cite{KouwenhofenReport,DeFranceshiReview,MarcKastner} on such systems mostly at low enough temperature with
negligible thermal and non-equilibrium fluctuations.

\begin{figure}
\resizebox{1.0\hsize}{!}{\includegraphics*{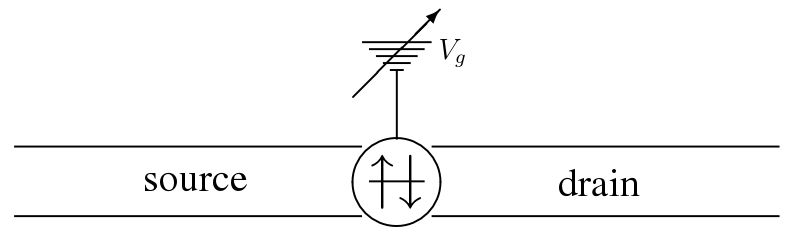}}
\caption{Set-up: single level quantum dot connected with two superconducting leads via coupling constants ${\rm \Gamma}_s$}
\label{figure1}
\end{figure}

\begin{figure}
\resizebox{1.0\hsize}{!}{\includegraphics{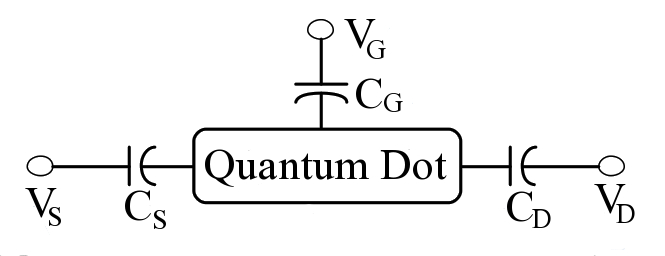}}
\caption{Equivalent capacitive circuit with coupling capacitances $C_{s}$, $C_{g}$ and $C_{d}$,
corresponding to the capacitances in
the source, gate and drain respectively.}
\label{figure2}
\end{figure}

All the systems mentioned above underlay  universal common features with the hybrid superconductor quantum dot devices we want to address
in this work \cite{DeFranceshiReview,SDattaBook2005,SDattacondmat2006,datta95} (i) Broadened energy levels  of the quantum-dot due to
hybridization with the leads. (ii) Spatial potential profile. (iii) A charging energy $U_{\rm 0}$ due to the potential profile. An insight
behind these issues have been highlighted recently \cite{AWGhost,Liang} for molecular dots. The device we study in this work is shown in
Figure \ref{figure1}.
It constitutes a spin degenerated quantum dot level, which is coupled to a pair
of biased superconductors contacts or leads (source and
drain). When a source-drain voltage $V_{d}$ is applied,
an electric current flows between the leads and across the quantum
dot. The biasing defines a  non-equilibrium steady state situation.
Such situation is coming from the frustration to establish simultaneously
an equilibrium  configuration with both leads under a given bias.
In addition, a gate voltage $V_{g}$ sets the quantum dot spectrum.
However, the charge energy can modify it whenever the density of states
is significant.
In response to the applied voltages, an actual potential develops inside
the dot, i.e., an effective electrostatic profile potential inside the
mesoscopic region exists in such a way, that it couples to  both the electronic non-equilibrium state
population and the non-equilibrium electric current.
That approach, as introduced by S. Datta \cite{SDattaBook2005}, links the
electrostatic profile to the electronic population of
the quantum dot \cite{SDattaBook2005,HighBiasCoulombBlockageSDatta} via the non-equilibrium
Keldysh formalism (NEGF) \cite{keldysh65,haug}.
The whole system is modeled by coupling capacitances which represents the drain, source
and gate contributions to the self-consistent electrostatic problem. Incoming electrons
have to overcome an energy barrier (Coulomb blockade).
On the other hand, gate or source-drain voltage can lower or increase this energy barrier.
These source, drain and gate electrodes capacitances (see Figure \ref{figure2})
constitute a simple capacitive model (in experiments \cite{DeFranceshiReview,RalphBlackTinkham},
these capacitances are measured) from which  $U_{\cal L}$, the Laplacian part of the potential, can be obtained.
In addition, the charge in dot can be expressed as the sum of the charges in
the coupling capacitances. It yields the Poisson contribution $U_{P}$, to the total
potential $U$, as a function of the dot population.
In other words, we solves the self-consistency (SC) of the total electrostatic potential
$U=U_{\cal L}+U_{P}$ together with the dot population. After that, the electric current is evaluated.

Previous to the self-consistent program, the non-equilibrium current through the dot and electronic occupation in the dot are worked out.
We emphasize that the calculation is carry out in a general framework. However, we confine our attention
to the single particle current contribution. We adapt the SC to two different approximation regimes.
In section (\ref{CapacitiveModel}), the equivalent capacitive circuit (Figure \ref{figure2}) is introduced, the spatial potential profile
$U$ is calculated within the capacitive model. The SC scheme is
applied for two cases \cite{resitance,resitance1}. First, the so called restricted case, where the gap is the bigger energy scale and the
coupling QD-Leads is
of the order of the charging energy (${\rm \Delta} \gg {{\rm \Gamma}}_{L,R}\simeq U$). In this case, quantitative results are expected to be
accurate.
We also make calculations for the so called unrestricted case, where the charging energy is the dominant energy scale
${\rm \Delta}\simeq U\gg {{\rm \Gamma}}_{L,R}$. In this case the results are quantitatively less accurate.
The experiments of Ralph et al \cite{RalphBlackTinkham} were done in this regime. Their ${\rm I}-V$ characteristic shows
that the spacing of the energy levels are subjected to strong fluctuations. According to our model, the fluctuations are due to complex
multilevel charging effects.
Our hybrid S/QD/S system has been studied in previous theoretical works \cite{KKang,MRoderoI,MRoderoII,JKoenig}.
However, to our knowledge, the coupled SC scheme which describe charging effects has not been considered so far.
This is an important step, then, gauge invariant independence of the results as well independence of the zero reference
voltage is fulfilled \cite{MBuettiker1,HongGuo1}. Our model use experimental values of the equivalent capacitances
\cite{SDattaBook2005}. To this respect, pioneering work done by Meir, Wingreen and Lee \cite{MeirI,MeirII} for a N/QD/N systems,
consider the interatomic Coulomb term $Un_{\uparrow}n_{\downarrow}$ as a measure  of the charging energy $e^{2}/C$. His purpose was to find
the main object of the non-equilibrium formalism, namely, the QD Green-Keldysh function, in which the influence of the leads on the QD is
taken into account. Due to the presence of the  Coulomb  term, its  equation of motion generates a two particle Green-Keldysh function. By
ignoring correlations with the leads, the equation of motion for the QD Green function closes after truncation of higher order equations
of motions. This solution (their Equation (8)), has two resonances, one at the energy level weighted by the probability that the other spin
degenerate level (raised by U) is vacant and another one at the energy level raised by $U$ weighted by the probability
that the level is occupied. It is correct for temperatures higher than the Kondo temperature and is exact in the non-interacting limit
($U=0$) and the isolated limit. Analogously, for a S/QD/S hybrid systems Kang \cite{KKang} has  obtained an expression for
the current through the QD (his Equation (8)), which is evaluated in the $U\to\infty$ limit (his Equation (13)). The QD Green function from
the very beginning does not contain off-diagonal terms that involves superconducting pairing, which excludes the possibility of Andreev
reflection processes. The presence in the equation for the current (his Equation (14)) of terms proportional to $(1-\langle
n_{-\sigma}\rangle)$ affects the contribution to the current of the considered level. In order to complete the outlined program one has to
calculate $\langle n_{-\sigma}\rangle$ self consistently which is not carried out. Instead, Kang calculate the current (his Equation (8))
where the spectral function is calculated  in the limit of zero coupling with the leads via a model taken from literature (his reference
[18]) and without taken into account the dependence of the contribution of one level to the current on the occupancy of the other. The point
of view which neglects the unavoidable influence of the bath (the leads) on the small system (the QD) is accomplished by factorizing the
density matrix ($\rho\left(t\right)=\rho_{QD}\bigotimes\rho_{Baths}$)
and integrating out the leads degrees of freedom which simplifies the Lioville-von Neumann equation (Equation 3.140 in
\cite{BreuerPetruccione}). This program is carried out by Kosov et al. for a S/QD/S system \cite{DKosov}. In this way, a Markovian
master equation is obtained and an expression for the current is calculated. In their Figure 2, they show the ${\rm I}-V$ characteristic of
a non-degenerated QD for a given set of parameters. In this case, the Cooper pair density in the QD is zero
\cite{BlackRalphTinkham}. For the sake of comparison, we restrict our calculations to this case. A similar but not identical approach was
done by Pfaller et al. \cite{Pfaller}. Also, the approach of both Kosov et al. and Pfaller et al. misses the energy
levels broadening as discussed in the introduction. This lack of broadening is a general deficit of Quantum Markov approach
\cite{EspositoGalperin}. In particular, Pfaller et al. \cite{Pfaller}, introduce a phenomenological broadening while our approach
derives it from first principles. In fact, within the Keldysh formalism, this broadening appears naturally (see our Equation (69) below).
Yeyati et al. \cite{MRoderoI} writes an expression for the current (his Equation (2) and Figure 2). They use that expression to explain the
experimental results of Ralph et al. \cite{RalphBlackTinkham}. Their calculation where done in the $U\to\infty$ limit. In addition, they
include charging effects, although they do not say explicitly in which way these effects are included.
In this respect, one has to realize that $U$ has important contributions from the QD mesoscopic charging effect. In the
$t\to -\infty$ the leads and the QD maintain independent thermal equilibrium, i.e., are uncoupled systems. When they become coupled the
Keldysh formalism yields the general behavior of the system. After a long enough time, this particular system reaches a steady state.

Our point of view is taken form the fact, that the charging of the QD is the origin of the Coulomb repulsion between two electron occupying
a two fold degenerate level. Therefore, we study the behavior of a noninteracting QD at $t\to-\infty$ where exact expressions are found. In
this way, we obtain a formally similar expression (Equation (72) below) for the current as Equation (12) in the work of Meir et al.
\cite{MeirI}. Later on, Coulomb repulsion is introduced via a self-consistent field (SCF) that depends dynamically on the applied bias
($H_{QD}+U_{SCF}$) and, in consequence, on the actual number of electrons in the QD. This approach constitutes the coupled Poisson NEGF
formalism that has been discussed in the context of molecular conductors by Datta \cite{SDattaBook2005,Liang}. We use a capacitive model in
section \ref{CapacitiveModel} to calculate $U_{SCF}$ and as discussed above, a numerical procedure is used to evaluate the current. Our
approach has the known disadvantage, of ignoring correlations in the QD (as pointed out in \cite{EspositoGalperin}). In that sense, there is
a proposal by Datta (Equation 3.4.9 in \cite{SDattaBook2005}) that improves the SCF method and permits more accurate quantitative results.
In section \ref{CapacitiveModel} we apply this improvement for the case when the Coulomb charging is greater than the value of the coupling
constants.
We discuss possible improvements of our approach in section \ref{conclusions}.

\section{Single Level QD-model: Derivation of Nonequilibrium Currents}
\label{Model} In macroscopic systems the task of deriving
 transport equations or generalized Ginzburg-Landau equations relies on quasi-classical Green functions \cite{NKopnin}.
 In addition, recently non-equilibrium transport in dirty Aluminium quasi one dimensional nanowires coupled with normal reservoirs  \cite{PekolaKlapwijk}
 was studied experimentally and theoretically with quasi-classical Green functions \cite{SnymanNazarov}.
As we want to include the possibility of  particle interference effects, we do no resort to such objects.
This point of view has been discussed in \cite{FClaro}.
Instead, we use the equation of motion method (EOM) technique of Keldysh formalism for generating
non-equilibrium states (see references \cite{Keldysh2003,RvaanLwewen,FPMarin,keldysh65}).
We consider a spin degenerated single orbital as a quantum dot connected
to superconductors leads. The hamiltonian which describes this system is
a generalized Anderson model \cite{PhysRev.124.41}. It reads
\begin{equation}
H = H_{S} + H_{QD} + H_{T},
\label{equation1}
\end{equation}
where $H_{S}$, $H_{QD}$ and $H_{T}$ stand for the superconducting leads,
the dot and the tunneling term, respectively.
$H_{S} =\displaystyle{\sum_{\eta}}H_{\eta}= H_{L} + H_{R}$
where $H_{L}$ and $H_{R}$ are the left and right lead Hamiltonians,
respectively. They are given, within the BCS model
\cite{PhysRev.108.1175}, by
\begin{equation}
H_{S}
=
\sum_{\eta\vec{k}\sigma}
\Psi^{\dagger}_{\eta\vec{k}\sigma}
\mathrm{H}^{0}_{\eta\vec{k}}
\Psi_{\eta\vec{k}\sigma},
\label{equation2}
\end{equation}
with
\begin{equation}
\mathrm{H}^{0}_{\eta\vec{k}}
=
\left(%
\begin{array}{cc}
\varepsilon_{\eta\vec{k}}&{\rm \Delta}_{\eta\vec{k}}
\\
{\rm \Delta}^{*}_{\eta\vec{k}}&-\varepsilon_{\eta\vec{k}}
\end{array}\right),
\label{equation3}
\end{equation}
where $\varepsilon_{\eta\vec{k}}$ is the conduction electron energy, ${\rm \Delta}_{\eta\vec{k}}$ is the superconductor gap, of the lead $\eta = L, R$. $\Psi^{\dagger}_{\eta\vec{k}\sigma}$ and
$\Psi_{\eta\vec{k}\sigma}$ are the Nambu spinors.
\begin{equation}
\Psi_{\eta\vec{k}\sigma}^{\dagger}
=
\left(%
a_{\eta\vec{k}\sigma}^{\dagger}
\quad
a_{\eta,-\vec{k},-\sigma}\right),
\qquad
\Psi_{\eta\vec{k}\sigma}
=
\left(%
\begin{array}{c}
a_{\eta\vec{k}\sigma}
\\
a^{\dagger}_{\eta,-\vec{k},-\sigma}
\end{array}\right).
\label{equation4}
\end{equation}
Here
$a^{\dagger}_{\eta\vec{k}\sigma}\left(a_{\eta\vec{k}\sigma}\right)$
denotes the creation (annihilation) operator for a conduction electron
with wave vector $\vec{k}$ and spin $\sigma$ in the $\eta = L, R$
superconductor lead.

$H_{QD}$ is the hamiltonian for the single-level quantum
dot  of energy ${\rm E}_{0}$:
\begin{equation}
H_{QD}
=
\sum_{\sigma}\phi^{\dagger}_{\sigma}\mathrm{H}^{QD}\phi_{\sigma},
\label{equation5}
\end{equation}
with
\begin{equation}
\mathrm{H}^{QD}
=
\left(%
\begin{array}{cc}
{\mathrm E}_{0}  & 0
\\
0 & -{\mathrm E}_{0}
\end{array}\right).
\label{equation6}
\end{equation}

The model  QD  does not contain the  Hubbard Coulomb repulsion interaction term. As explained in the introduction,
Coulomb repulsion is modeled by means of the inclusion of capacitances, which are taken independent of the charge in the QD.
The model also ignores possible superconducting correlations in the QD. For sufficiently small QDs, the discreteness of the
single energy levels suppress these correlations \cite{BlackRalphTinkham}.
The position of the energy level will be treated first as fixed by the gate potential with respect to the left lead, while the
effect of the applied voltage is taking into account by the coupled Poisson scheme.
The tunneling hamiltonian $H_{T}$ is given by:
\begin{equation}
H_{T}
=
\sum_{\eta\vec{k}\sigma}
\Psi_{\eta\vec{k}\sigma}^{\dagger}
\mathrm{H}^{I}_{\eta\vec{k}}\phi_{\sigma},
\label{equation7}
\end{equation}
with
\begin{equation}
\mathrm{H}^{I}_{\eta\vec{k}}
=
\left(%
\begin{array}{cc}
V_{\eta\vec{k}} & 0
\\
0 & -V_{\eta\vec{k}}
\end{array}\right).
\label{equation8}
\end{equation}
$H_{T}$ connects the dot to the biased superconducting leads and it
allows the electric charge flow. $V_{\eta\vec{k}}$ is the hybridization
matrix element between a conduction electron in the $\eta = L, R$ superconductor lead and
a localized electron on the dot with energy ${\rm E}_{0}$.
$\phi_{\sigma}^{\dagger}$ and $\phi_{\sigma}$ are the dot spinors
\begin{equation}
\phi_{\sigma}^{\dagger}
=
\left(d_{\sigma}^{\dagger}\quad d_{-\sigma}\right),
\qquad
\phi_{\sigma}
=
\left(%
\begin{array}{c}
d_{\sigma}
\\
d^{\dagger}_{-\sigma}
\end{array}\right),
\label{equation9}
\end{equation}
here, $d^{\dagger}_{\sigma}\left(d_{\sigma}\right)$ is
the creation (annihilation) operator for an electron on the dot.

The flow of electric charge from the terminal $\eta$ is given by
\begin{equation}
{\rm I}_{\eta}\left( t\right)
=
\left(-e\right)
\left\lbrack
-\frac{d\left\langle N_{\eta}\left(t\right)\right\rangle}{dt}
\right\rbrack
=
\frac{{\rm i} e}{\hbar}
\left\langle
\left\lbrack H_{T}\left(t\right), N_{\eta}\left(t\right)\right\rbrack
\right\rangle,
\label{equation10}
\end{equation}
where $-{\rm e}$ is the electron charge.
$\left\langle\cdots\right\rangle$ is the thermodynamical average over
the biased $L$ and $R$ leads at the temperature $T$, taken at time $t_{0}\rightarrow -\infty$, as indicated in the
Keldysh contour in appendix A.
\begin{equation}
\left\langle\cdots\right\rangle\equiv{Tr(\rho(t_{0})...)},\hspace{1cm}
\rho(t_{0})\equiv\frac{e^{-\beta(H-{\mu}N)}}{Tr(e^{-\beta(H-{\mu}N)})},
\label{equation11}
\end{equation}
and
$N_{\eta}=a^{\dagger}_{\eta\vec{k}\sigma}a_{\eta\vec{k}\sigma}$ is
the ``number of particle'' operator. Book keeping calculations using Equation (\ref{equation10}), leads to
\begin{equation}
{\rm I}_{\eta}\left(t\right)
=
\frac{2e}{\hbar}V_{\eta}\Re\sum_{\vec{k}\sigma}
{\rm F}^{<}_{\eta\vec{k}\sigma}\left(t, t\right).
\label{equation12}
\end{equation}
${\rm F}^{<}_{\eta\vec{k}\sigma}\left(t, t'\right)={\rm i}\langle d^{\dagger}_{\sigma}\left(t'\right)a_{\eta\vec{k}\sigma}\left(t\right)\rangle$ is the lesser
Keldysh Green function,
\begin{eqnarray}
{\rm F}_{\eta\vec{k}\sigma}\left(t, t'\right)
&\equiv &
-{\rm i}\langle%
{\rm T_{{\rm {K}}}}a_{\eta\vec{k}\sigma}\left(t\right)
           d^{\dagger}_{\sigma}\left(t'\right)\rangle
\nonumber\\
&\equiv &
-{\rm i}{\rm\Theta}\left(t,t'\right)\langle
a_{\eta\vec{k}\sigma}\left(t\right)
           d^{\dagger}_{\sigma}\left(t'\right)\rangle
+{\rm i}{\rm\Theta}\left(t',t\right)\langle
d^{\dagger}_{\sigma}\left(t'\right)a_{\eta\vec{k}\sigma}\left(t\right)
           \rangle
\nonumber\\
&\equiv &
{\rm\Theta}\left(t,t'\right){\rm F}^{>}_{\eta\vec{k}\sigma}\left(t, t'\right)
+{\rm\Theta}\left(t',t\right){\rm F}^{<}_{\eta\vec{k}\sigma}\left(t, t'\right),
\label{equation13}
\end{eqnarray}
and ${\rm T_{K}}$ is the time-ordering operator, the action of which is  to rearrange product of  operators, such that operator  with later times, on the Keldysh contour are placed to the left of the product.
 Hereafter, for simplicity, we replace
$V_{\eta\vec{k}}$  by an average~$V_{\eta}$ at the Fermi surfaces( $V_{\eta\vec{k}}\equiv\sqrt{\langle|V_{\eta\vec{k}}|^{2}\rangle_{FS}}$)  of the
leads $L$ and $R$.
Using the scheme given in appendix A for the rate of change of Equation (\ref{equation13}), we proceed to obtain the equation of motion:
\begin{equation}
{\rm i}\frac{\partial {\rm F}_{\eta\vec{k}\sigma}\left(t, t'\right)}{\partial t}
=
\delta\left(t,t'\right)\left\langle\left\{ a_{\eta\vec{k}\sigma}\left(t\right),d^{\dagger}_{\sigma}\left(t\right)\right\}\right\rangle
-
{\rm i}\left\langle{\rm T_{{\rm {K}}}}\left\lbrack a_{\eta\vec{k}\sigma}\left(t\right),H\right\rbrack d^{\dagger}_{\sigma}\left(t'\right)\right\rangle.
\label{equation14}
\end{equation}
Which leads to
\begin{equation}
\left({\rm i}\frac{\partial}{\partial t}
      -
      \epsilon_{\eta\vec{k}}\right)
{\rm F}_{\eta\vec{k}\sigma}\left(t, t'\right)
=
-\sigma{\rm \Delta}_{\eta}{\mathcal F}_{\eta\vec{k}\sigma}\left(t, t'\right)
+
V_{\eta}{\rm G}_{\sigma}\left(t, t'\right),
\label{equation15}
\end{equation}
where
\begin{eqnarray}
{\mathcal F}_{\eta\vec{k}\sigma}\left(t, t'\right)
& = &
-{\rm i}
\left\langle
{\rm T_{{\rm {K}}}}
a^{\dagger}_{\eta\vec{k},-\sigma}\left(t\right)
d^{\dagger}_{\sigma}\left(t'\right)
\right\rangle,
\label{equation16}
\\
{\rm G}_{\sigma}\left(t, t'\right)
& = &
-{\rm i}
\left\langle
{\rm T_{{\rm {K}}}}d_{\sigma}\left(t\right)
           d^{\dagger}_{\sigma}\left(t'\right)
\right\rangle.
\label{equation17}
\end{eqnarray}
Note that ${\rm G}_{\sigma}\left(t, t'\right)$ is the QD single particle Green's function.
Similarly, ${\mathcal F}_{\eta\vec{k}\sigma}\left(t, t'\right)$
satisfies the equation of motion:
\begin{equation}
\left(%
{\rm i}\frac{\partial}{\partial t}
       +
       \epsilon_{\eta\vec{k}}\right)
{\mathcal F}_{\eta\vec{k}\sigma}\left(t, t'\right)
=
-\sigma{\rm \Delta}_{\eta}{\rm F}_{\eta\vec{k}\sigma}\left(t, t'\right)
-
V_{\eta}{\mathcal G}_{\sigma}\left(t, t'\right),
\label{equation18}
\end{equation}
where
\begin{equation}
{\mathcal G}_{\sigma}\left(t, t'\right)
=
-{\rm i}\left\langle
{\rm T_{{\rm {K}}}}d^{\dagger}_{-\sigma}\left(t\right)
           d^{\dagger}_{\sigma}\left(t'\right)
\right\rangle.
\label{equation19}
\end{equation}
Here ${\mathcal G}_{\sigma}\left(t, t'\right)$ is the
QD of two-particle Green's function.

Equations (\ref{equation15}) and (\ref{equation18}) can be written in a compact form
as follows (see Appendix B):
\begin{equation}
\left(%
\begin{array}{cc}
\displaystyle{{\rm i}\frac{\partial}{\partial t}} - \epsilon_{\eta\vec{k}}
&
\sigma{\rm \Delta}_{\eta}
\\
&\\
\sigma{\rm \Delta}_{\eta}
&
\displaystyle{{\rm i}\frac{\partial}{\partial t}} + \epsilon_{\eta\vec{k}}
\end{array}\right)
\left(%
\begin{array}{cc}
{\rm F}_{\eta\vec{k}\sigma}(t,t')&\quad\widetilde{\mathcal F}_{\eta\vec{k}\sigma}(t,t')\\
&\\
&\\
{\mathcal F}_{\eta\vec{k}\sigma}(t,t')&\quad\widetilde{\rm F}_{\eta\vec{k}\sigma}(t,t')
\end{array}\right)
=
V_{\eta}\sigma_{z}\left(%
\begin{array}{cc}
{\rm G}_{\sigma}\left(t, t'\right)&\quad\widetilde{\mathcal G}_{\eta\vec{k}\sigma}(t,t')\\
&\\
&\\
{\mathcal G}_{\sigma}\left(t, t'\right)&\quad\widetilde{\rm G}_{\eta\vec{k}\sigma}(t,t') \end{array}\right).
\label{equation20}
\end{equation}
We introduce the  tilde Keldysh-Green functions:
\begin{eqnarray}
\widetilde{\mathcal F}_{\eta\vec{k}\sigma}\left(t, t'\right)&=&-{\rm i}\left\langle
{\rm T_{{\rm {K}}}}a_{\eta\vec{k}\sigma}\left(t\right)
           d_{-\sigma}\left(t'\right)
\right\rangle,
\nonumber\\
\quad\widetilde{\rm F}_{\eta\vec{k}\sigma}\left(t, t'\right)&=&-{\rm i}\left\langle
{\rm T_{{\rm {K}}}}a^{\dagger}_{\eta-\vec{k}-\sigma}\left(t\right)
           d_{-\sigma}\left(t'\right)\right\rangle,
\nonumber\\
\widetilde{\mathcal G}_{\sigma}\left(t, t'\right)&=&-{\rm i}\left\langle
{\rm T_{{\rm {K}}}}d_{\sigma}\left(t\right)
           d_{-\sigma}\left(t'\right)
\right\rangle,
\nonumber\\
\quad\widetilde{\rm G}_{\sigma}\left(t, t'\right)&=&-{\rm i}\left\langle
{\rm T_{{\rm {K}}}}d^{\dagger}_{-\sigma}\left(t\right)
           d_{-\sigma}\left(t'\right)
\right\rangle.
\label{equation21}
\end{eqnarray}
Consider the following $2\times2$ matrix whose elements are the unperturbed Green-Keldysh functions, i.e., defined for $V_{\eta}=0$,
\begin{equation}
\begin{array}{c|l}
\begin{array}{cl}
\left(
\begin{array}{cc}
{\rm g}_{\eta\vec{k}\sigma}\left(t,t'\right)&\quad{\rm f}_{\eta\vec{k}\sigma}\left(t,t'\right)\\
&\\
\widetilde{{\rm f}}_{\eta\vec{k}\sigma}\left(t,t'\right)&\quad\widetilde{{\rm g}}_{\eta\vec{k}\sigma}\left(t,t'\right)\\
\end{array}
\right)
\end{array}
\quad\hbox{where}\quad&
\begin{array}{l}
{\rm g}_{\eta\vec{k}\sigma}\left(t,t'\right)\equiv-{\rm i}\left\langle{\rm T_{{\rm {K}}}}a_{\eta\vec{k}\sigma}\left(t\right)a^{\dagger}_{\eta\vec{k}\sigma}\left(t'\right)\right\rangle_{0}\\
\\
{\rm f}_{\eta\vec{k}\sigma}\left(t,t'\right)\equiv-{\rm i}\left\langle{\rm T_{{\rm {K}}}}a_{\eta\vec{k}\sigma}\left(t\right)a_{\eta-\vec{k}-\sigma}\left(t'\right)\right\rangle_{0}\\
\\
\widetilde{\rm f}_{\eta\vec{k}\sigma}\left(t,t'\right)\equiv-{\rm i}\left\langle{\rm T_{{\rm {K}}}}a^{\dagger}_{\eta-\vec{k}-\sigma}\left(t\right)a^{\dagger}_{\eta\vec{k}\sigma}\left(t'\right)\right\rangle_{0}\\
\\
\widetilde{\rm g}_{\eta\vec{k}\sigma}\left(t,t'\right)\equiv-{\rm i}\left\langle{\rm T_{{\rm {K}}}}a^{\dagger}_{\eta-\vec{k}-\sigma}\left(t\right)a_{\eta-\vec{k}-\sigma}\left(t'\right)\right\rangle_{0}\
\end{array}
\end{array}
\label{equation22}
\end{equation}

According to appendix A, their equations of motions are given by:
\begin{eqnarray}
\left({\rm i}\frac{\partial}{\partial t}-\epsilon_{\eta\vec{k}}\right){\rm g}_{\eta\vec{k}\sigma}\left(t,t'\right)+\sigma{\rm \Delta}_{\eta}\widetilde{\rm f}_{\eta\vec{k}\sigma}\left(t,t'\right)&=&\delta\left(t,t'\right),\\
\left({\rm i}\frac{\partial}{\partial t}-\epsilon_{\eta\vec{k}}\right){\rm f}_{\eta\vec{k}\sigma}\left(t,t'\right)+\sigma{\rm \Delta}_{\eta}\widetilde{\rm g}_{\eta\vec{k}\sigma}\left(t,t'\right)&=&0,\\
\left({\rm i}\frac{\partial}{\partial t}+\epsilon_{\eta\vec{k}}\right)\widetilde{\rm f}_{\eta\vec{k}\sigma}\left(t,t'\right)+\sigma{\rm \Delta}_{\eta}{\rm g}_{\eta\vec{k}\sigma}\left(t,t'\right)&=&0,\\
\left({\rm i}\frac{\partial}{\partial t}+\epsilon_{\eta\vec{k}}\right)\widetilde{\rm g}_{\eta\vec{k}\sigma}\left(t,t'\right)+\sigma{\rm \Delta}_{\eta}{\rm f}_{\eta\vec{k}\sigma}\left(t,t'\right)&=&\delta\left(t,t'\right).
\label{equations23-26}
\end{eqnarray}
These equations can be written in matrix form as follows:
\begin{equation}
\begin{array}{cl}
\left(
\begin{array}{cc}
\displaystyle{{\rm i}\frac{\partial}{\partial t}}-\epsilon_{\eta\vec{k}}&\sigma{\rm \Delta}_{\eta}\\
&\\
\sigma{\rm \Delta}_{\eta}&\displaystyle{{\rm i}\frac{\partial}{\partial t}}+\epsilon_{\eta\vec{k}}\\
\end{array}
\right)
\left(
\begin{array}{cc}
{\rm g}_{\eta\vec{k}\sigma}\left(t,t'\right)&\quad{\rm f}_{\eta\vec{k}\sigma}\left(t,t'\right)\\
&\\
&\\
\widetilde{{\rm f}}_{\eta\vec{k}\sigma}\left(t,t'\right)&\quad\widetilde{{\rm g}}_{\eta\vec{k}\sigma}\left(t,t'\right)\\
\end{array}
\right)
\end{array}
=
\left(
\begin{array}{cc}
\delta\left(t,t'\right)&0\\
&\\
&\\
0&\delta\left(t,t'\right)\\
\end{array}
\right).
\label{equation27}
\end{equation}

The Equation (\ref{equation20}) can be written as an integral along the Keldysh contour
${\rm C_{K}}$, (for an explanation see  Appendix B).
\begin{eqnarray}
\left(%
\begin{array}{cc}
{\rm F}_{\eta\vec{k}\sigma}(t,t')&\quad\widetilde{\mathcal F}_{\eta\vec{k}\sigma}(t,t')\\
&\\
{\mathcal F}_{\eta\vec{k}\sigma}(t,t')&\quad\widetilde{\rm F}_{\eta\vec{k}\sigma}(t,t')
\end{array}\right)
=
\int_{\rm C_{K}}{\rm d}t''
\left(%
\begin{array}{cc}
{\rm g}_{\eta\vec{k}\sigma}\left(t, t''\right)
&
\widetilde{\rm f}_{\eta\vec{k}\sigma}\left(t, t''\right)\\
&\\
{\rm f}_{\eta\vec{k}\sigma}\left(t, t''\right)
&
\widetilde{\rm {g}}_{\eta\vec{k}\sigma}\left(t, t''\right)
\end{array}\right)\times
\nonumber\\
V_{\eta}\sigma_{z}
\left(%
\begin{array}{cc}
{\rm G}_{\sigma}\left(t'', t'\right)&\quad\widetilde{{\mathcal G}}_{\sigma}\left(t'', t'\right)\\
&\\
{\mathcal G}_{\sigma}\left(t'', t'\right)&\quad\widetilde{{\rm G}}_{\sigma}\left(t'', t'\right)\\
\end{array}\right).
\label{equation28}
\end{eqnarray}

From the last expression  one can read for ${\rm F}_{\eta\vec{k}\sigma}(t,t')$ the equation:
\begin{eqnarray}
{\rm F}_{\eta\vec{k}\sigma}(t,t')
=
V_{\eta}\int_{{\rm C_{K}}} {\rm d}t''\, \left[ {\rm g}_{\eta\vec{k}\sigma}(t,t''){\rm G}_{\sigma}(t'',t')\right.
-\nonumber\\
\left. \widetilde{\rm f}_{\eta\vec{k}\sigma}(t,t''){\mathcal G}_{\sigma}(t'',t')\right].
\label{equation29}
\end{eqnarray}

We now apply the procedure explained in appendix C, in order to obtain the ${\rm F}_{\eta\vec{k}\sigma}(t,t')$ lesser component, we obtain:
\begin{eqnarray}
{\rm F}^{<}_{\eta\vec{k}\sigma}(t,t')
=
V_{\eta}&&\left\{\int^{\infty}_{-\infty} {\rm d}t''\, \left\lbrack {\rm g}^{\left(\rm r\right)}_{\eta\vec{k}\sigma}(t,t''){\rm G}^{<}_{\sigma}(t'',t')\right.\right.
-
\left. \widetilde{\rm f}^{\left(\rm r\right)}_{\eta\vec{k}\sigma}(t,t''){\mathcal G}^{<}_{\sigma}(t'',t')\right\rbrack
+\nonumber\\
&&\left.\int^{\infty}_{-\infty} {\rm d}t''\, \left\lbrack {\rm g}^{<}_{\eta\vec{k}\sigma}(t,t''){\rm G}^{\left(\rm a\right)}_{\sigma}(t'',t')\right.
-
\left. \widetilde{\rm f}^{<}_{\eta\vec{k}\sigma}(t,t''){\mathcal G}^{\left(\rm a\right)}_{\sigma}(t'',t')\right\rbrack\right\}.
\label{equation30}
\end{eqnarray}
Furthermore, the superscripts ${}^{(<),(>)}{}^{\left({\rm r}\right),\left({\rm a}\right)}$ correspond to lesser, greater, retarded, advanced  Green's functions respectively.

Therefore, from Equation (\ref{equation12}), ${\rm I}_{\eta}(t)$, can be written as:
\begin{equation}
{\rm I}_{\eta}(t)={\rm I}^{(1)}_{\eta}(t)+{\rm I}^{(2)}_{\eta}(t),
\label{equation31}
\end{equation}
with
\begin{eqnarray}
{\rm I}^{(1)}_{\eta}(t)
=
\frac{2e}{\hbar}\Re\sum_{\sigma}\int^{\infty}_{-\infty}{\rm d}t'
\left\{\left\lbrack V^{2}_{\eta} \sum_{\vec{k}}{\rm g}^{\left(\rm r\right)}_{\eta\vec{k}\sigma}(t,t')\right\rbrack{\rm G}^{<}_{\sigma}(t',t)\right.
+\nonumber\\
\left.\left\lbrack V^{2}_{\eta}\sum_{\vec{k}}{\rm g}^{<}_{\eta\vec{k}\sigma}(t,t')\right\rbrack{\rm G}^{\left({\rm a}\right)}_{\sigma}(t',t)\right\},
\label{equation32}
\end{eqnarray}
\begin{eqnarray}
{\rm I}^{(2)}_{\eta}(t)
=
-\frac{2e}{\hbar}\Re\sum_{\sigma}\int^{\infty}_{-\infty}{\rm d}t'
\left\{\left\lbrack V^{2}_{\eta} \sum_{\vec{k}}\widetilde{\rm f}^{\left(\rm r\right)}_{\eta\vec{k}\sigma}(t,t')\right\rbrack{\mathcal G}^{<}_{\sigma}(t',t)\right.
+\nonumber\\
\left.\left\lbrack V^{2}_{\eta}\sum_{\vec{k}}\widetilde{\rm f}^{<}_{\eta\vec{k}\sigma}(t,t')\right\rbrack{\mathcal G}^{\left({\rm a}\right)}_{\sigma}(t',t)\right\}.
\label{equation33}
\end{eqnarray}

When applying the Fourier transformations, the Equations (\ref{equation32}) and (\ref{equation33}) can be expressed as
\begin{eqnarray}
{\rm I}^{(1)}_{\eta}(t)
&=&
\frac{2e}{h}\Re\sum_{\sigma}\int^{\infty}_{-\infty}{\rm d}\omega\int^{\infty}_{-\infty}\frac{{\rm d}\omega'}{{2\pi}}{\rm e}^{-{\rm i}(\omega - \omega')t}\times
\nonumber
\\
&&\left\lbrack{\rm \Sigma}^{\left({\rm r}\right)}_{\eta}(\omega){\rm G}^{<}_{\sigma}(\omega,\omega') +  {\rm \Sigma}^{<}_{\eta}(\omega){\rm G}^{\left({\rm a}\right)}_{\sigma}(\omega,\omega')\right\rbrack,
\label{equation34}
\end{eqnarray}
and
\begin{eqnarray}
{\rm I}^{(2)}_{\eta}(t)
&=&
-\frac{2e}{h}\Re\sum_{\sigma}\left\{{\rm e}^{-2{\rm i}\mu_{\eta}t}\int^{\infty}_{-\infty}{\rm d}\omega\int^{\infty}_{-\infty}\frac{{\rm d}\omega'}{{2\pi}}{\rm e}^{-{\rm i}(\omega - \omega')t}\times\right.
\nonumber
\\
&&\left.\left\lbrack\widetilde{\rm \Xi}^{\left({\rm r}\right)}_{\eta}(\omega)\sigma{\mathcal G}^{<}_{\sigma}(\omega,\omega') +\widetilde{\rm \Xi}^{<}_{\eta}(\omega)\sigma{\mathcal G}^{\left({\rm a}\right)}_{\sigma}(\omega,\omega')\right\rbrack\right\},
\label{equation35}
\end{eqnarray}
with
\begin{eqnarray}
V^{2}_{\eta}\sum_{\vec{k}}{\rm g}^{\left({\rm r}\right)}_{\eta\vec{k}\sigma}\left(t,t'\right)&\equiv &
\int^{\infty}_{-\infty}\frac{{\rm d}\omega}{2\pi}{\rm e}^{-{\rm i}\omega\left(t-t'\right)}{\rm \Sigma}^{\left({\rm r}\right)}_{\eta}\left(\omega\right),
\\
V^{2}_{\eta}\sum_{\vec{k}}{\rm g}^{<}_{\eta\vec{k}\sigma}\left(t,t'\right)&\equiv &
\int^{\infty}_{-\infty}\frac{{\rm d}\omega}{2\pi}{\rm e}^{-{\rm i}\omega\left(t-t'\right)}
{\rm \Sigma}^{<}_{\eta}\left(\omega\right),
\\
V^{2}_{\eta}\sum_{\vec{k}}\widetilde{\rm f}^{({\rm r})}_{\eta\vec{k}\sigma}\left(t,t'\right)&\equiv&
\int^{\infty}_{-\infty}{\rm e}^{-2{\rm i}\mu_{\eta}t}\sigma\frac{{\rm d}\omega}{2\pi}{\rm e}^{-{\rm i}\omega\left(t-t'\right)}\widetilde{\rm \Xi}^{({\rm r})}_{\eta}\left(\omega\right),
\\
V^{2}_{\eta}\sum_{\vec{k}}\widetilde{\rm f}^{<}_{\eta\vec{k}\sigma}\left(t,t'\right)&\equiv&
\int^{\infty}_{-\infty}{\rm e}^{-2{\rm i}\mu_{\eta}t}\sigma\frac{{\rm d}\omega}{2\pi}{\rm e}^{-{\rm i}\omega\left(t-t'\right)}\widetilde{\rm \Xi}^{<}_{\eta}\left(\omega\right).
\label{equations36}
\end{eqnarray}

In appendixes  D to G we  evaluate the unperturbed Green's functions ${{\rm g}}^{\left({\rm r}\right)}_{\eta\vec{k}\sigma}\left(t,t'\right)$,\\ ${\rm g}^{<}_{\eta\vec{k}\sigma}\left(t,t'\right)$,
$\widetilde{{\rm f}}^{\left({\rm r}\right)}_{\eta\vec{k}\sigma}\left(t,t'\right)$ and $\widetilde{{\rm f}}^{<}_{\eta\vec{k}\sigma}\left(t,t'\right)$ in the wide band limit.

We summarize these results:
\begin{eqnarray}
{\rm \Sigma}^{\left({\rm r}\right)}_{\eta}(\omega)
&=&
-{{\rm \Gamma}}_{\eta}\left\lbrack\frac{\omega-\mu_{\eta}}{{\rm \Delta}_{\eta}}\zeta({\rm \Delta}_{\eta},\omega-\mu_{\eta})
+
{\rm i}\zeta(\omega-\mu_{\eta},{\rm \Delta}_{\eta})\right\rbrack,
\nonumber
\\
{\rm \Sigma}^{<}_{\eta}(\omega)
&=&
2{\rm i}{{\rm \Gamma}}_{\eta}\zeta(\omega-\mu_{\eta},{\rm \Delta}_{\eta}){\rm f}(\omega-\mu_{\eta}),
\nonumber
\\
\widetilde{{\rm \Xi}}^{\left({\rm r}\right)}_{\eta}(\omega)
&=&
{{\rm \Gamma}}_{\eta}
\left\lbrack\zeta({\rm \Delta}_{\eta},\omega+\mu_{\eta})
+
{\rm i}\frac{{\rm \Delta}_{\eta}}{\omega+\mu_{\eta}}\zeta(\omega+\mu_{\eta},{\rm \Delta}_{\eta})\right\rbrack,
\\
\widetilde{{\rm \Xi}}^{<}_{\eta}(\omega)
&=&
-2{\rm i}{{\rm \Gamma}}_{\eta}\frac{{\rm \Delta}_{\eta}}{\omega+\mu_{\eta}}\zeta(\omega+\mu_{\eta},{\rm \Delta}_{\eta}){\rm f}(\omega+\mu_{\eta}),
\nonumber
\\
\zeta(\omega,\omega')
&\equiv&
{\rm \Theta}(|\omega|-|\omega'|)\frac{|\omega|}{\sqrt{\omega^{2}-\omega'^{2}}}.
\nonumber
\label{equatio37}
\end{eqnarray}

All these expressions will used below.

\section{QD Green Function.}
We need to evaluate the  most important objet for calculations, namely the QD Green's functions given by Equation (\ref{equation17})
and Equation (\ref{equation19}), as well their respective tilde functions:
\begin{eqnarray}
\widetilde{{\rm G}}_{\sigma}\left(t, t'\right)
&=&
-{\rm i}\left\langle
{\rm T_{{\rm {K}}}}d^{\dagger}_{-\sigma}\left(t\right)
           d_{-\sigma}\left(t'\right)
\right\rangle,\nonumber\\
\widetilde{{\mathcal G}}_{\sigma}\left(t, t'\right)
&=&
-{\rm i}\left\langle
{\rm T_{{\rm {K}}}}d_{\sigma}\left(t\right)
           d_{-\sigma}\left(t'\right)
\right\rangle.
\end{eqnarray}

Again using the scheme given in Appendix A,  their  equation of motion are:

\begin{eqnarray}
\begin{array}{ll}
{\rm i}\displaystyle{\frac{\partial}{\partial t}}
\left(
\begin{array}{cc}
{\rm G}_{\sigma}\left(t,t'\right)&\quad\widetilde{\mathcal G}_{\sigma}\left(t,t'\right)\\
&\\
&\\
{\mathcal G}_{\sigma}\left(t,t'\right)&\quad\widetilde{\rm G}_{\sigma}\left(t,t'\right)\\
\end{array}
\right)
=\nonumber\\
\nonumber\\
\left(
\begin{array}{cc}
\delta\left(t,t'\right)-{\rm i}\left\langle{\rm T_{{\rm {K}}}}\lbrack d_{\sigma}\left(t\right),H\rbrack d^{\dagger}_{\sigma}\left(t'\right)\right\rangle &-{\rm i}\left\langle{\rm T_{{\rm {K}}}}\lbrack d_{\sigma}\left(t\right),H\rbrack d_{-\sigma}\left(t'\right)\right\rangle\\
&\\
&\\
-{\rm i}\left\langle{\rm T_{{\rm {K}}}}\lbrack d^{\dagger}_{-\sigma}\left(t\right),H\rbrack d^{\dagger}_{\sigma}\left(t'\right)\right\rangle & \delta\left(t,t'\right)-{\rm i}\left\langle{\rm T_{{\rm {K}}}}\lbrack d^{\dagger}_{-\sigma}\left(t\right),H\rbrack d_{-\sigma}\left(t'\right)\right\rangle
\end{array}
\right).
\end{array}
\label{equation38}
\end{eqnarray}
Which develops to:
\begin{eqnarray}
{\rm i}\displaystyle{\frac{\partial}{\partial t}}
{\rm G}_{\sigma}\left(t,t'\right)
&=&
\delta\left(t,t'\right)-{\rm i}{\rm E}_{0}\left\langle{\rm T_{{\rm K}}} d_{\sigma}\left(t\right)d^{\dagger}_{\sigma}\left(t'\right)\right\rangle
-{\rm i}\displaystyle{\sum_{\eta\vec{k}}}V_{\eta}\left\langle {\rm T_{{\rm K}}}a_{\eta\vec{k}\sigma}\left(t\right)d^{\dagger}_{\sigma}\left(t'\right)\right\rangle\nonumber\\
&=&
\delta\left(t,t'\right)+{\rm E}_{0}{\rm G}_{\sigma}\left(t,t'\right)+\displaystyle{\sum_{\eta\vec{k}}}V_{\eta}{\rm F}_{\eta\vec{k}\sigma}\left(t,t'\right),\\
{\rm i}\displaystyle{\frac{\partial}{\partial t}}{\mathcal G}_{\sigma}\left(t,t'\right)
&=&
{\rm i}{\rm E}_{0}\left\langle{\rm T_{{\rm K}}} d^{\dagger}_{-\sigma}\left(t\right)d^{\dagger}_{\sigma}\left(t'\right)\right\rangle
+{\rm i}\displaystyle{\sum_{\eta\vec{k}}}V_{\eta}\left\langle {\rm T_{{\rm K}}}a^{\dagger}_{\eta -\vec{k}-\sigma}\left(t\right)d^{\dagger}_{\sigma}\left(t'\right)\right\rangle\nonumber\\
&=&
-{\rm E}_{0}{\mathcal{G}}_{\sigma}\left(t,t'\right)-\displaystyle{\sum_{\eta\vec{k}}}V_{\eta}{\mathcal{F}}_{\eta\vec{k}\sigma}\left(t,t'\right),
\\
{\rm i}\displaystyle{\frac{\partial}{\partial t}}\widetilde{\mathcal G}_{\sigma}\left(t,t'\right)
&=&
-{\rm i}{\rm E}_{0}\left\langle{\rm T_{{\rm K}}} d_{\sigma}\left(t\right)d_{-\sigma}\left(t'\right)\right\rangle
-{\rm i}\displaystyle{\sum_{\eta\vec{k}}}V_{\eta}\left\langle {\rm T_{{\rm K}}}a_{\eta\vec{k}\sigma}\left(t\right)d_{-\sigma}\left(t'\right)\right\rangle\nonumber
\\
&=&
{\rm E}_{0}\widetilde{\mathcal G}_{\sigma}\left(t,t'\right)+\displaystyle{\sum_{\eta\vec{k}}}V_{\eta}\widetilde{\mathcal F}_{\eta\vec{k}\sigma}\left(t,t'\right),
\\
{\rm i}\displaystyle{\frac{\partial}{\partial t}}
\widetilde{\rm G}_{\sigma}\left(t,t'\right)
&=&
\delta\left(t,t'\right)+{\rm i}{\rm E}_{0}\left\langle{\rm T_{{\rm K}}} d^{\dagger}_{-\sigma}\left(t\right)d_{-\sigma}\left(t'\right)\right\rangle
+{\rm i}\displaystyle{\sum_{\eta\vec{k}}}V_{\eta}\left\langle {\rm T_{{\rm K}}}a^{\dagger}_{\eta\vec{k}\sigma}\left(t\right)d_{-\sigma}\left(t'\right)\right\rangle\nonumber
\\
&=&
\delta\left(t,t'\right)-{\rm E}_{0}\widetilde{\rm G}_{\sigma}\left(t,t'\right)-\displaystyle{\sum_{\eta\vec{k}}}V_{\eta}\widetilde{\mathcal F}_{\eta\vec{k}\sigma}\left(t,t'\right).
\label{equation39}
\end{eqnarray}
This can be written as:
\begin{eqnarray}
\left(%
\begin{array}{cc}
{\rm i}\displaystyle{\frac{\partial}{\partial t}} - {\rm E}_{0}
&
0
\\
&\\
0
&
{\rm i}\displaystyle{\frac{\partial}{\partial t}} + {\rm E}_{0}
\end{array}\right)
\left(%
\begin{array}{cc}
{\rm G}_{\sigma}\left(t, t'\right)&\quad\widetilde{\mathcal G}_{\sigma}\left(t, t'\right)\\
&\\
&\\
{\mathcal G}_{\sigma}\left(t, t'\right)&\quad\widetilde{\rm G}_{\sigma}\left(t, t'\right)
\end{array}\right)
=
\left(%
\begin{array}{cc}
\delta\left(t, t'\right)&0\\
&\\
&\\
0&\delta\left(t, t'\right)
\end{array}\right)
\nonumber\\
+
\sum_{\eta\vec{k}}V_{\eta}\sigma_{z}\left(%
\begin{array}{cc}
{\rm F}_{\eta\vec{k}\sigma}\left(t, t'\right)&\quad\widetilde{\mathcal F}_{\eta\vec{k}\sigma}\left(t, t'\right)\\
&\\
&\\
{\mathcal F}_{\eta\vec{k}\sigma}\left(t, t'\right)&\quad\widetilde{\rm F}_{\eta\vec{k}\sigma}\left(t, t'\right) \end{array}\right).
\label{equation40}
\end{eqnarray}
When $V_{\eta}=0$ one has:
\begin{equation}
\left(%
\begin{array}{cc}
{\rm i}\displaystyle{\frac{\partial}{\partial t}} - {\rm E}_{0}
&
0
\\
&\\
0
&
{\rm i}\displaystyle{\frac{\partial}{\partial t}} + {\rm E}_{0}
\end{array}\right)
\left(%
\begin{array}{cc}
{\rm G}_{0}\left(t,t'\right)&0\\
&\\
&\\
0&\widetilde{\rm G}_{0}\left(t,t'\right)
\end{array}\right)
=
\left(%
\begin{array}{cc}
\delta\left(t,t'\right)&0\\
&\\
&\\
0&\delta\left(t,t'\right)
\end{array}\right),
\label{equation41}
\end{equation}
with:
\begin{eqnarray}
{\rm G}_{0}\left(t,t'\right)=&-{\rm i}\left\langle{\rm T_{{\rm K}}}d_{\sigma}\left(t\right)d^{\dagger}_{\sigma}\left(t'\right)\right\rangle_{0},&\qquad\widetilde{\rm G}_{0}\left(t,t'\right)=-{\rm i}\left\langle{\rm T_{{\rm K}}}d^{\dagger}_{-\sigma}\left(t\right)d_{-\sigma}\left(t'\right)\right\rangle_{0}.\nonumber\\
{\rm G}_{0}\left(t,t'\right)
\equiv &
{\rm G}_{\sigma}\left(t,t'\right)\arrowvert_{V_{\eta}=0},&\qquad
\widetilde{\rm G}_{0}\left(t,t'\right)
\equiv
\widetilde{\rm G}_{\sigma}\left(t,t'\right)\arrowvert_{V_{\eta}=0}.
\label{equation42}
\end{eqnarray}
\newpage
The last two equations can be written as:
\begin{eqnarray}
\left(%
\begin{array}{cc}
{\rm i}\displaystyle{\frac{\partial}{\partial t}} - {\rm E}_{0}
&
0
\\
&\\
0
&
{\rm i}\displaystyle{\frac{\partial}{\partial t}} + {\rm  E}_{0}
\end{array}\right)
\left(%
\begin{array}{cc}
{\rm G}_{\sigma}\left(t, t'\right)-{\rm G}_{0}\left(t, t'\right)&\quad\widetilde{\mathcal G}_{\sigma}\left(t, t'\right)\\
&\\
&\\
{\mathcal G}_{\sigma}\left(t, t'\right)&\quad\widetilde{\rm G}_{\sigma}\left(t, t'\right)-\widetilde{\rm G}_{0}\left(t, t'\right)
\end{array}\right)
=\nonumber\\
\sum_{\eta\vec{k}}V_{\eta}\sigma_{z}\left(%
\begin{array}{cc}
{\rm F}_{\eta\vec{k}\sigma}\left(t, t'\right)&\widetilde{\mathcal F}_{\eta\vec{k}\sigma}\left(t, t'\right)\\
&\\
&\\
{\mathcal F}_{\eta\vec{k}\sigma}\left(t, t'\right)&\widetilde{\rm F}_{\eta\vec{k}\sigma}\left(t, t'\right) \end{array}\right).\nonumber\\
\label{equation43}
\end{eqnarray}

We write the last equation in its equivalent convolution integral along the Keldysh contour (see Appendix B):
\begin{eqnarray}
\left(%
\begin{array}{cc}
{\rm G}_{\sigma}(t,t')-{\rm G}_{0}(t,t')&\widetilde{\mathcal G}_{\sigma}(t,t')\\
&\\
&\\
{\mathcal G}_{\sigma}(t,t')&\widetilde{\rm G}_{\sigma}(t,t')-\widetilde{\rm G}_{0}(t,t')
\end{array}\right)
=
\displaystyle{\int_{\rm C_{K}}{\rm d}t''}
\left(%
\begin{array}{cc}
{\rm G}_{0}(t,t'')
&
0
\\
\\
0
&
\widetilde{\rm G}_{0}(t,t'')
\end{array}\right)
\times\nonumber\\
\sum_{\eta\vec{k}}V_{\eta}\sigma_{z}
\left(
\begin{array}{cc}
{\rm F}_{\eta\vec{k}\sigma}\left(t'', t'\right)&\quad\widetilde{\mathcal F}_{\eta\vec{k}\sigma}\left(t'', t'\right)\\
&\\
&\\
{\mathcal F}_{\eta\vec{k}\sigma}\left(t'', t'\right)&\quad\widetilde{\rm F}_{\eta\vec{k}\sigma}\left(t'', t'\right)
\end{array}\right).\nonumber\\
\label{equation44}
\end{eqnarray}

An equivalent way to write the last equation (using equation (\ref{equation27})) as a convolution
of ${\sf \Sigma}_{\sigma}\left(t,t'\right)$ and
${\sf G}_{\sigma}\left(t,t'\right)$ is:
\begin{equation}
{\sf G}_{\sigma}\left(t,t'\right)
=
{\sf G}_{0}\left(t,t'\right)
+\int_{{\rm C_{K}}}{\rm d}t''{\sf G}_{0}\left(t,t'\right){\sf \Sigma}_{\sigma}\left(t',t''\right){\sf G}_{\sigma}\left(t'',t'\right),
\label{equation45}
\end{equation}
with:
\begin{equation}
\begin{array}{cl}
{\sf G}_{\sigma}\left(t,t'\right)&
\equiv
\left(
\begin{array}{cl}
{\rm G}_{\sigma}\left(t,t'\right)&\quad\widetilde{\mathcal G}_{\sigma}\left(t,t'\right)\\
&\\
&\\
{\mathcal G}_{\sigma}\left(t,t'\right)&\quad\widetilde{\rm G}_{\sigma}\left(t,t'\right)
\end{array}\right),
\qquad{\sf G}_{0}\left(t,t'\right)
\equiv
\left(
\begin{array}{cc}
{\rm G}_{0}\left(t,t'\right)&0\\
&\\
&\\
0&\widetilde{\rm G}_{0}\left(t,t'\right)
\end{array}
\right).
\end{array}
\label{equation46}
\end{equation}
and
\begin{equation}
\begin{array}{ll}
{\sf \Sigma}_{\sigma}\left(t,t'\right)
\equiv
\displaystyle{\int_{{\rm C_{K}}}}{\rm d}t''\left(
\begin{array}{cc}
V^{2}_{\eta}\displaystyle{\sum_{\eta\vec{k}}}{\rm g}_{\eta\vec{k}\sigma}\left(t'',t'\right)&-V^{2}_{\eta}\displaystyle{\sum_{\eta\vec{k}}}\widetilde{\rm f}_{\eta\vec{k}\sigma}\left(t'',t'\right)\\
&\\
-V^{2}_{\eta}\displaystyle{\sum_{\eta\vec{k}}}{\rm f}_{\eta\vec{k}\sigma}\left(t'',t'\right)&V^{2}_{\eta}\displaystyle{\sum_{\eta\vec{k}}}\widetilde{{\rm g}}_{\eta\vec{k}\sigma}\left(t'',t'\right)\\
\end{array}
\right)
\end{array}
\label{equation47}
\end{equation}

We are interested  in two regimes: A first regime in which $U_{\rm 0}\sim{\rm \Gamma}<{\rm \Delta}$ and the Coulomb blockade effects is neglected because in this case the couplings to the leads are not extremely small and the dot capacitance is large enough. A second regime for $U_{\rm 0}\sim{\rm \Delta}>{\rm \Gamma}$ where Coulomb blockade effects must be taken into account. For both regimes and from now on, we are interested in the case $eV>{\rm \Delta}$, where multiple Andreev reflection \cite{Andreev1} processes  is strongly suppressed. Therefore only the single particle current ($SP$) have to be considered $\rm{I}_{SP}$. From the above considerations we have that the Keldysh Green function ${\mathcal G}_{\sigma}\left(\omega\right)$, which carries information of the  quantum dot two-particle Green's function can be neglected and all relevant information is contained in
${\rm G}_{\sigma}\left(\omega\right)$.

The Keldysh Green function becomes spin independent,  ${\rm G}_{\sigma}\left(\omega\right)\equiv{\rm G}\left(\omega\right)$. The element 11 of Equation (\ref{equation45}) is given by:
\begin{equation}
{\rm G}\left(t,t'\right)
=
{\rm G}_{0}\left(t,t'\right)
+
\int_{{\rm C_{K}}}{\rm d}t''\int_{{\rm C_{K}}}{\rm d}t'''\,\,{\rm G}_{0}\left(t,t''\right){\rm \Sigma}\left(t'',t'''\right){\rm G}\left(t''',t'\right).
\label{equation50}
\end{equation}

Again, using the recipe given in appendix C, we obtain for ${\rm G}^{<}\left(t,t'\right)$ and ${\rm G}^{\left(\rm a\right)}\left(t,t'\right)$:
\begin{eqnarray}
{\rm G}^{<}\left(t,t'\right)
=
{\rm G}_{0}^{<}\left(t,t'\right)+\left[\int^{\infty}_{-\infty}{\rm d}t''\int^{\infty}_{-\infty}{\rm d}t'''\right.&&\left.{\rm G}^{\left(\rm r\right)}_{0}\left(t,t''\right){\rm \Sigma}^{\left(\rm r\right)}\left(t'',t'''\right){\rm G}^{<}\left(t''',t'\right)+\right.
\nonumber\\
&&\left.{\rm G}^{\left(\rm r\right)}_{0}\left(t,t''\right){\rm \Sigma}^{<}\left(t'',t'''\right){\rm G}^{\left(\rm a\right)}\left(t''',t'\right)\right.+
\nonumber\\
&&\left.{\rm G}^{<}_{0}\left(t,t''\right){\rm \Sigma}^{\left(\rm a\right)}\left(t'',t'''\right){\rm G}^{\left(\rm a\right)}\left(t''',t'\right)\right].
\label{equation51}
\end{eqnarray}
\begin{equation}
{\rm G}^{\left(\rm a\right)}\left(t,t'\right)
=
{\rm G}_{0}^{\left(\rm a\right)}\left(t,t'\right)+\int^{\infty}_{-\infty}{\rm d}t''\int^{\infty}_{-\infty}{\rm d}t'''\,\,{\rm G}^{\left(\rm a\right)}_{0}\left(t,t''\right){\rm \Sigma}^{\left(\rm a\right)}\left(t'',t'''\right){\rm G}^{\left(\rm a\right)}\left(t''',t'\right).
\label{equation52}
\end{equation}
Taking the Fourier transform of Equations (\ref{equation51}) and
(\ref{equation52}), results in a set of algebraic equations:
\begin{eqnarray}
{\rm G}^{<}\left(\omega,\omega'\right)
&=&
2\pi\delta\left(\omega-\omega'\right){\rm G}_{0}^{<}\left(\omega\right)
+
{\rm G}_{0}^{\left(\rm r\right)}\left(\omega\right){\rm \Sigma}^{\left(\rm r\right)}\left(\omega\right){\rm G}^{<}\left(\omega,\omega'\right)
+
\nonumber\\
&&{\rm G}_{0}^{\left(\rm r\right)}\left(\omega\right){\rm \Sigma}^{<}\left(\omega\right){\rm G}^{\left(\rm a\right)}\left(\omega,\omega'\right)+
{\rm G}_{0}^{<}\left(\omega\right){\rm \Sigma}^{\left(\rm a\right)}\left(\omega\right){\rm G}^{\left(\rm a\right)}\left(\omega,\omega'\right).
\label{equation53}
\\
\nonumber\\
{\rm G}^{\left(\rm a\right)}\left(\omega,\omega'\right)
&=&
2\pi\delta\left(\omega-\omega'\right){\rm G}_{0}^{\left(\rm a\right)}\left(\omega\right)
+
{\rm G}_{0}^{\left(\rm a\right)}\left(\omega\right){\rm \Sigma}^{\left(\rm a\right)}\left(\omega\right){\rm G}^{\left(\rm a\right)}\left(\omega,\omega'\right).
\label{equation54}
\end{eqnarray}
Dot Keldysh Green's functions
${\rm G}_{\sigma}^{<}\left(\omega,\omega'\right)$ and
${\rm G}_{\sigma}^{{\rm\left(a\right)}}\left(\omega,\omega'\right)$
are  below straightforward evaluated. In this regime  quantities such as  currents are independent of time.
Therefore, we have:
\begin{eqnarray*}
{\rm G}^{<}\left(\omega,\omega'\right)
& = &
2\pi\delta\left(\omega - \omega'\right){\rm G}^{<}\left(\omega\right),
\\
{\rm G}^{{\rm\left(a\right)}}\left(\omega,\omega'\right)
& = &
2\pi\delta\left(\omega - \omega'\right)
{\rm G}^{{\rm\left(a\right)}}\left(\omega\right).
\label{equation55}
\end{eqnarray*}
Therefore the Equations (\ref{equation53}) and (\ref{equation54}) result
\begin{eqnarray}
{\rm G}^{<}\left(\omega\right)
&=&
{\rm G}_{0}^{<}\left(\omega\right)
+
{\rm G}_{0}^{\left(\rm r\right)}\left(\omega\right){\rm \Sigma}^{\left(\rm r\right)}\left(\omega\right){\rm G}^{<}\left(\omega\right)
+
\nonumber\\
&&{\rm G}_{0}^{\left(\rm r\right)}\left(\omega\right){\rm \Sigma}^{<}\left(\omega\right){\rm G}^{\left(\rm a\right)}\left(\omega\right)+
{\rm G}_{0}^{<}\left(\omega\right){\rm \Sigma}^{\left(\rm a\right)}\left(\omega\right){\rm G}^{\left(\rm a\right)}\left(\omega\right).
\label{equation56}
\\
\nonumber\\
{\rm G}^{\left(\rm a\right)}\left(\omega\right)
&=&
{\rm G}_{0}^{\left(\rm a\right)}\left(\omega\right)
+
{\rm G}_{0}^{\left(\rm a\right)}\left(\omega\right){\rm \Sigma}^{\left(\rm a\right)}\left(\omega\right){\rm G}^{\left(\rm a\right)}\left(\omega\right).
\label{equation57}
\end{eqnarray}
Solving (\ref{equation57}),
\begin{equation}
\mathrm{G}^{\left(\rm a\right)}\left(\omega\right)
=
\frac{1}{{\mathrm{G}^{\left(\rm a\right)}_{0}\left(\omega\right)}^{-1}-{\rm \Sigma}^{\left(\rm a\right)}\left(\omega\right)}
=
\frac{1}{\omega-{\rm E}_{0}-{\rm \Sigma}^{\left(\rm a\right)}\left(\omega\right)}
=
\mathrm{G}^{\left(\rm r\right)}\left(\omega\right)^{*}.
\label{equation58}
\end{equation}
Moreover, we know that
\begin{eqnarray}
{\rm G}_{0}^{<}\left(\omega\right)&\propto&\delta\left(\omega-{\rm E_{0}}\right)\quad\hbox{and}\\
{\rm G}_{0}^{\left(\rm a\right)}\left(\omega\right)&=&\left(\omega-{\rm E_{0}}-{\rm i}0^{+}\right)^{-1},
\label{equation59}
\end{eqnarray}
resulting
\begin{equation}
{\rm G}_{0}^{<}\left(\omega\right){\rm \Sigma}^{\left(\rm a\right)}\left(\omega\right){\rm G}^{\left(\rm a\right)}\left(\omega\right)
=
-{\rm G}_{0}^{<}\left(\omega\right).
\label{equation60}
\end{equation}
The Equation (\ref{equation56}) is reduced to
\begin{equation}
{\rm G}^{<}\left(\omega\right)
=
{\rm G}_{0}^{\left(\rm r\right)}\left(\omega\right){\rm \Sigma}^{\left(\rm r\right)}\left(\omega\right){\rm G}^{<}\left(\omega\right)
+
{\rm G}_{0}^{\left(\rm r\right)}\left(\omega\right){\rm \Sigma}^{<}\left(\omega\right){\rm G}^{\left(\rm a\right)}\left(\omega\right),
\label{equation61}
\end{equation}
\begin{eqnarray}
{\rm G}^{<}\left(\omega\right)
&=&
\frac{{\rm \Sigma}^{<}\left(\omega\right){\rm G}^{\left(\rm a\right)}\left(\omega\right)}{{\rm G}_{0}^{\left(\rm r\right)}\left(\omega\right)^{-1}-{\rm \Sigma}^{\left(\rm r\right)}\left(\omega\right)}
=
{\rm \Sigma}^{<}\left(\omega\right)\arrowvert{\rm G}^{\left(\rm r\right)}\left(\omega\right)\arrowvert^{2}\nonumber\\
&=&
\pi{\rm \Sigma}^{<}\left(\omega\right)\frac{-\Im {\rm G}^{\left(\rm r\right)}\left(\omega\right)/\pi}{\Im\left({\rm G}^{\left(\rm r\right)}\left(\omega\right)\right)^{-1}}
=
\pi\frac{{\rm \Sigma}^{<}\left(\omega\right)}{-\Im {\rm \Sigma}^{\left(\rm r\right)}\left(\omega\right)}\rho\left(\omega\right).
\label{equation62}
\end{eqnarray}
Here $\rho\left(\omega\right)$ is the so-called quantum dot spectral
function which is given in terms of the imaginary part ($\Im$) of the retarded
Keldysh Green function $\mathrm{G^{\left(\rm r\right)}}\left(\omega\right)$,
\begin{equation}
\rho\left(\omega\right)
 =
-\,\frac{1}{\pi}
\Im{\rm G}^{\rm\left(r\right)}\left(\omega\right)
=
-\frac{1}{\pi}\frac{\Im{\rm \Sigma}^{\left(\rm r\right)}\left(\omega\right)}{\left\lbrack\omega-{\rm E}_{0}-\Re{\rm \Sigma}^{\left(\rm r\right)}\left(\omega\right)\right\rbrack^{2}+\left\lbrack\Im{\rm \Sigma}^{\left(\rm r\right)}\left(\omega\right)\right\rbrack^{2}}.
\label{equation63}
\end{equation}
From Equation (\ref{equation32}) the single particle current (${\rm I}_{SP}$) results in:
\begin{equation}
{\rm I}_{\eta SP}\left(V,{\rm E}_{0}\right)
 =
\frac{4e}{h}\Re
\int_{-\infty}^{\infty}{\rm d}\omega
\left\lbrack%
{\rm \Sigma}_{\eta}^{\rm\left(r\right)}\left(\omega\right)
{\rm G}^{<}\left(\omega\right)
+
{\rm \Sigma}_{\eta}^{<}\left(\omega\right)
{\rm G}^{\rm\left(a\right)}\left(\omega\right)
\right\rbrack .
\label{equation64}
\end{equation}
Substituting (\ref{equation58}) and (\ref{equation62}) in (\ref{equation64})
\begin{eqnarray}
{\rm I}_{\eta SP}\left(V,{\rm E}_{0}\right)
&=&
\frac{4e}{h}
\int_{-\infty}^{\infty}{\rm d}\omega
\left\lbrack%
\pi\Im{\rm \Sigma}_{\eta}^{\rm\left(r\right)}\left(\omega\right)
\frac{\Im{\rm \Sigma}^{<}\left(\omega\right)}
{\Im {\rm \Sigma}^{\left(\rm r\right)}\left(\omega\right)}\rho\left(\omega\right)
+
\Im{\rm \Sigma}_{\eta}^{<}\left(\omega\right)
\Im{\rm G}^{\rm\left(r\right)}\left(\omega\right)
\right\rbrack \nonumber\\
&=&
\frac{4\pi e}{h}
\int_{-\infty}^{\infty}{\rm d}\omega\rho\left(\omega\right)
\left\lbrack%
\Im{\rm \Sigma}_{\eta}^{\rm\left(r\right)}\left(\omega\right)
\frac{\Im{\rm \Sigma}^{<}\left(\omega\right)}
{\Im {\rm \Sigma}^{\rm\left(r\right)}\left(\omega\right)}\rho\left(\omega\right)
-
\Im{\rm \Sigma}_{\eta}^{<}\left(\omega\right)
\right\rbrack
\nonumber\\
&=&
\frac{4\pi e}{h}
\int_{-\infty}^{\infty}{\rm d}\omega\frac{\rho\left(\omega\right)}{\rm \Gamma\left(\omega\right)}
\left\lbrack%
{\rm \Gamma}_{\eta}\left(\omega\right)\Im{\rm \Sigma}^{<}\left(\omega\right)
-
{\rm \Gamma}\left(\omega\right)\Im{\rm \Sigma}_{\eta}^{<}\left(\omega\right)
\right\rbrack .
\label{equation65}
\end{eqnarray}
with
$
{\rm \Gamma}_{\eta}\left(\omega\right)
=
-\Im{\rm \Sigma}_{\eta}^{\rm\left(r\right)}\left(\omega\right)
=
{{\rm \Gamma}}_{\eta}\zeta\left(\omega,{\rm \Delta}_{\eta}\right)
$
and ${\rm \Gamma}\left(\omega\right)=\sum_{\eta}{\rm \Gamma}_{\eta}\left(\omega\right)
$.
In our regime, $eV>{\rm \Delta}$, therefore, $\Re{\rm \Sigma}^{\left(\rm r\right)}\left(\omega\right)$ in the above equations is zero.
We use the expression for ${\rm \Sigma}^{\left(\rm r\right)}\left(\omega\right)$ from appendix D, and obtain the single particle current
${\rm I}_{SP} \equiv \left({\rm I}_{R,SP} - {\rm I}_{L,SP}\right)/2$:
\begin{equation}
{\rm I}_{SP}\left(V,{\rm E}_{0}\right)
=
\frac{8\pi e}{h}
\int^{\infty}_{-\infty}{\rm d}\omega\,
\frac{{\rm \Gamma}_{L}\left(\omega\right){\rm \Gamma}_{R}\left(\omega+eV\right)}
      {{\rm \Gamma}_{L}\left(\omega\right)
       +
      {\rm \Gamma}_{R}\left(\omega+ eV\right)}\,
\rho\left(\omega\right)
\left\lbrack
{\rm f}\left(\omega\right) - {\rm f}\left(\omega+eV\right)
\right\rbrack .
\label{equation66}
\end{equation}
$-eV=\mu_{L}-\mu_{R}$
correspond to the applied voltage between the superconductors
electrodes with chemical potential $\mu_{\eta}$. In the
following, we fix the chemical potential $\mu_{L}=0$
and use $eV$ as a measure of $\mu_{R}$. In addition, the QD energy $E_{0}$ is measure with respect to $\mu_{L}$.
On the other hand, the limits of integration are given by the functions ${\rm \Gamma}_{L}\left(\omega\right)$
and ${\rm \Gamma}_{R}\left(\omega+ eV\right)$.
The extra $2\pi$ factor arises from the dot
Keldysh Green functions. $\rho\left(\omega\right)$ and ${\rm \Gamma}\left(\omega\right)$
are given by
\begin{eqnarray}
\rho\left(\omega\right)
&=&
\frac{{\rm \Gamma}\left(\omega\right)/\pi}
     {\left(\omega - {\rm E}_{0}\right)^{2}
      +
      {{\rm \Gamma}}^{2}\left(\omega\right)},
\label{equation67a}\\
{{\rm \Gamma}}\left(\omega\right)
& = &
{{\rm \Gamma}}_{L}\left(\omega\right)
+
{{\rm \Gamma}}_{R}\left(\omega+ eV\right).
\label{equation67b}
\end{eqnarray}

At steady state there is no net flow into or out of
the mesoscopic channel or quantum dot which yields a stationary particle number in it.
The population number $N$, at the dot, is given by
\begin{equation}
N
=
2\left\lbrack -{\rm i}{\rm G}^{<}\left(t,t\right)\right\rbrack
=
2\int^{\infty}_{-\infty}\frac{{\rm d}\omega}{2\pi{\rm i}}\,
{\rm G}^{<}\left(\omega\right),
\label{equation68}
\end{equation}
which becomes a weighted average over the $L$ and $R$ contacts
\begin{equation}
N
=
2\int^{\infty}_{-\infty}{\rm d}\omega\,\rho\left(\omega\right)
\left\lbrack
\frac{{\rm \Gamma}_{L}\left(\omega\right)}{{\rm \Gamma}\left(\omega\right)}\,
{\rm f}\left(\omega\right)
+
\frac{{\rm \Gamma}_{R}\left(\omega + eV\right)}{{\rm \Gamma}\left(\omega\right)}\,
{\rm f}\left(\omega+eV\right)
\right\rbrack.
\label{equation69}
\end{equation}
For the N/QD/N case, $\Gamma_{R,L}$  are just constants. This case was study in the context of  the generalized quantum master approach
(section IV in \cite{EspositoGalperin}). That approach permits the inclusion of broadening in a natural way. They obtained Equations similar
to our Equations (\ref{equation66})-(\ref{equation69}).
%
\section{\label{CapacitiveModel}Coupled Poisson-Nonequilibrum Green Function Scheme: The Capacitive Model}

So far, we are not included the side effects of a potential profile
inside the mesoscopic channel. On the one hand, its inclusion takes in order zero or Hartree approximation the electron-electron interaction in the QD. Its inclusion also guarantees current independence  from the choice of zero potential \cite{HongGuo1}. Such potential is induced by the action of source, drain and gate applied voltages. In principle, we have to
couple the number population equation Equation (\ref{equation68}), with electric field $U$. However,  since the number of quantum
levels in the channel is small the particle number variation is
negligible, the potential profile variation inside the
channel is negligible. Then it is appropriate visualize the channel as an
equivalent circuit framework (Figure \ref{figure2}). In this framework we associate capacitances $C_{d}$, $C_{s}$ and $C_{g}$ to the drain, source
and gate, respectively. Whenever drain, source and gate bias potentials
$V_{d}$, $V_{s}$ and $V_{g}$, respectively, are present, there is an electrostatic potential $V_{QD}$ inside the QD,  it induces an energy shift of the QD energy level
$U = -e\left(V_{QD} - V_{0}\right)$, $V_{0}$ are channel electrostatic
potential before we apply the source and drain biases,
respectively.

The electronic population before and after we apply the
biases mentioned above are given by
\begin{eqnarray}
-eN_{0}
& = &
C_{d}V_{0} + C_{s}V_{0} + C_{g}V_{0},
\\
-eN
& = &
C_{d}\left(V_{QD} - V_{d}\right) + C_{s}\left(V_{QD} - V_{s}\right)
+
C_{g}\left(V_{QD} - V_{g}\right),
\label{equation70}
\end{eqnarray}
respectively. It leads us to
\begin{equation}
-e\Delta N
\equiv
-e\left(N - N_{0}\right)
=
C_{E}\left(V_{QD} - V_{0}\right) - C_{d}V_{d} - C_{s}V_{s} - C_{g}V_{g},
\label{equation71}
\end{equation}
where $C_{E} = C_{d} +  C_{s} + C_{g}$. Therefore, the energy shift $U$ is given by:
\begin{equation}
U = U_{\cal L} + \frac{e^{2}}{C_{E}}\,\Delta N,
\label{equation72}
\end{equation}
and where
\begin{equation}
U_{\cal L}
\equiv
\frac{C_{d}}{C_{E}}\left(-eV_{d}\right)
+
\frac{C_{s}}{C_{E}}\left(-eV_{s}\right)
+
\frac{C_{g}}{C_{E}}\left(-eV_{g}\right).
\label{equation73}
\end{equation}
In the the expression for $U$, $U_{\cal L}$ represents a uniform shift for all levels, whereas the second term
(the Poisson contribution denoted $U_{P}$ in the introduction) represents a level
repulsion which is proportional to the averaged occupation of the QD level refereed to $N_{0}$, and proportional  to the charging energy $U_{\rm 0}=e^{2}/C_{E}$.

On the other hand, one has $\Delta N$, from Equations (\ref{equation68}) and (\ref{equation71}) given by:
\begin{equation}
\Delta N
=
2\int^{\infty}_{-\infty}\frac{{\rm d}\omega}{2\pi{\rm i}}\,
\left\lbrack{\rm G}^{<}\left(\omega,U\right)-{\rm G}^{<}\left(\omega,-eV_{0}\right)\right\rbrack .
\label{equation74}
\end{equation}
In the expression for ${\rm G}^{<}\left(\omega,U\right)$ (Equation (\ref{equation62})), the energy level shifts only, ($E_{0}\Rightarrow(E_{0}+U)$)
in the expression for  the QD spectral function $\rho\left(\omega - U\right)$.
Equations (\ref{equation72}) and (\ref{equation74}) are coupled non-linear equations with unknowns $U$ and $\Delta N$. We solve  the coupled equations via an iteration procedure.
First we guest a value for $\Delta N$, plug this value in $U$,  then we calculate $\Delta N$ with the equation:
\begin{equation}
\Delta N
=
2\int^{\infty}_{-\infty}{\rm d}\omega\,\rho\left(\omega - U\right)\,
\frac{{\rm \Gamma}_{L}\left(\omega\right)
      {\rm f}\left(\omega\right)
      +
      {\rm \Gamma}_{R}\left(\omega+eV\right){\rm f}\left(\omega+eV\right)}
      {{\rm \Gamma}_{L}\left(\omega\right)
      +
      {\rm \Gamma}_{R}\left(\omega+eV\right)},
\label{equation75}
\end{equation}
and so on until convergence is achieved. With the final value of $U$ obtained for a given bias  voltage $V$, $I_{SP}$ is calculated via the equation:
\begin{equation}
{\rm I}_{SP}\left(V,U\right)
=
\frac{8\pi e}{h}
\int^{\infty}_{-\infty}{\rm d}\omega\,
\frac{{\rm \Gamma}_{L}\left(\omega\right){\rm \Gamma}_{R}\left(\omega+eV\right)}
     {{\rm \Gamma}_{L}\left(\omega\right)
      +
      {\rm \Gamma}_{R}\left(\omega+eV\right)}\,
\rho\left(\omega -U\right)
\left\lbrack
{\rm f}\left(\omega\right) - {\rm f}\left(\omega+eV\right)
\right\rbrack .
\label{equation76}
\end{equation}

In summary, the procedure for computing $\rm{I}$
consists of the following steps.
i) Determine the spectral density.
ii) Specify $V_{g}$, $V_{d}$ and $V_{s}$ and coupling constants.
iii) Iteratively solve (\ref{equation75}) and (\ref{equation72}).
iv) Evaluate the current from (\ref{equation76}) for the $V_{g}$, $V_{d}$ and $V_{s}$.
Once a converged $U$ has been found, the current is finally evaluated.

The way we consider electron-electron interactions, imposes restrictions  on the possible values of the charging energy $U_{\rm 0}$.
For the self-consistent scheme to be valid, we have to assume, that ${\rm \Delta} \gg {\rm \Gamma}_{L,R}\simeq U_{\rm 0}$.
However, less precisely quantitative results, although qualitative correct results can be obtained if ${\rm \Delta}\simeq U_{0}\gg {{\rm \Gamma}}_{L,R}$,
when the  called Coulomb Blockade energy dominates over the coupling constants. For this case, we use the improvement of the SCF method discussed in the introduction \cite{SDattaBook2005,resitance,resitance1}. The self consistent generalizes to:
\begin{eqnarray}
U_{\uparrow} = U_{\cal L} + \frac{e^{2}}{C_{E}}\,(N_{\downarrow}-N_{0}),
\label{equation77a}\\
U_{\downarrow} = U_{\cal L} - \frac{e^{2}}{C_{E}}\,(N_{\uparrow}-N_{0}).
\label{equation77b}
\end{eqnarray}
where the up-spin level feels a potential due to the down-spin electrons and viceversa.
Notice the different signs, which reflects the Coulomb repulsion between otherwise degenerate levels.
\begin{equation}
N_{\uparrow}
=
\int^{\infty}_{-\infty}{\rm d}\omega\,\rho\left(\omega - U_{\uparrow}\right)\,
\frac{{\rm \Gamma}_{L}\left(\omega\right)
      {\rm f}\left(\omega\right)
      +
      {\rm \Gamma}_{R}\left(\omega+eV\right){\rm f}\left(\omega+eV\right)}
      {{\rm \Gamma}_{L}\left(\omega\right)
      +
      {\rm \Gamma}_{R}\left(\omega+eV\right)},
\label{equation78}
\end{equation}
\begin{equation}
N_{\downarrow}
=
\int^{\infty}_{-\infty}{\rm d}\omega\,\rho\left(\omega - U_{\downarrow}\right)\,
\frac{{\rm \Gamma}_{L}\left(\omega\right)
      {\rm f}\left(\omega\right)
      +
      {\rm \Gamma}_{R}\left(\omega+eV\right){\rm f}\left(\omega+eV\right)}
      {{\rm \Gamma}_{L}\left(\omega\right)
      +
      {\rm \Gamma}_{R}\left(\omega+eV\right)}.
\label{equation79}
\end{equation}

\begin{equation}
N
=
N_{\uparrow}
+
N_{\downarrow}.
\label{equation80}
\end{equation}
Here, $N_{\uparrow}$ and $N_{\downarrow}$ are the population of the spin-up and spin-down levels.
Once the values of $U_{\uparrow}$ and $U_{\downarrow}$ are calculated, ${\rm I}_{SP}$ is calculated from:

\begin{eqnarray}
{\rm I}_{SP}(V,U_{\uparrow},U_{\downarrow})
=
\frac{4\pi e}{h}
\int^{\infty}_{-\infty}{\rm d}\omega\,
\frac{{\rm \Gamma}_{L}\left(\omega\right){\rm \Gamma}_{R}\left(\omega+eV\right)}
     {{\rm \Gamma}_{L}\left(\omega\right)
      +
      {\rm \Gamma}_{R}\left(\omega+eV\right)}
\Bigl\{\rho\left(\omega -U_{\uparrow}\right)+\rho\left(\omega -U_{\downarrow}\right)\Bigr\}
\times\nonumber\\
\Bigl\lbrack
{\rm f}\left(\omega\right) - {\rm f}\left(\omega+eV\right)
\Bigr\rbrack .
\label{equation81}
\end{eqnarray}
As Datta has pointed out \cite{SDattaBook2005}, the approach described above (called unrestricted SCF) can lead to a better  quantitative agreement in comparison with a conceptually correct multi-level Master equation calculation.

\section{\label{ResandDisc}Numerical Results and Remarks on Experiments.}
\subsection{First case: ${\rm \Delta} \gg {{\rm \Gamma}}_{L,R}\simeq U_{\rm 0}$}

\begin{figure}
\resizebox{1.0\hsize}{!}{\includegraphics*{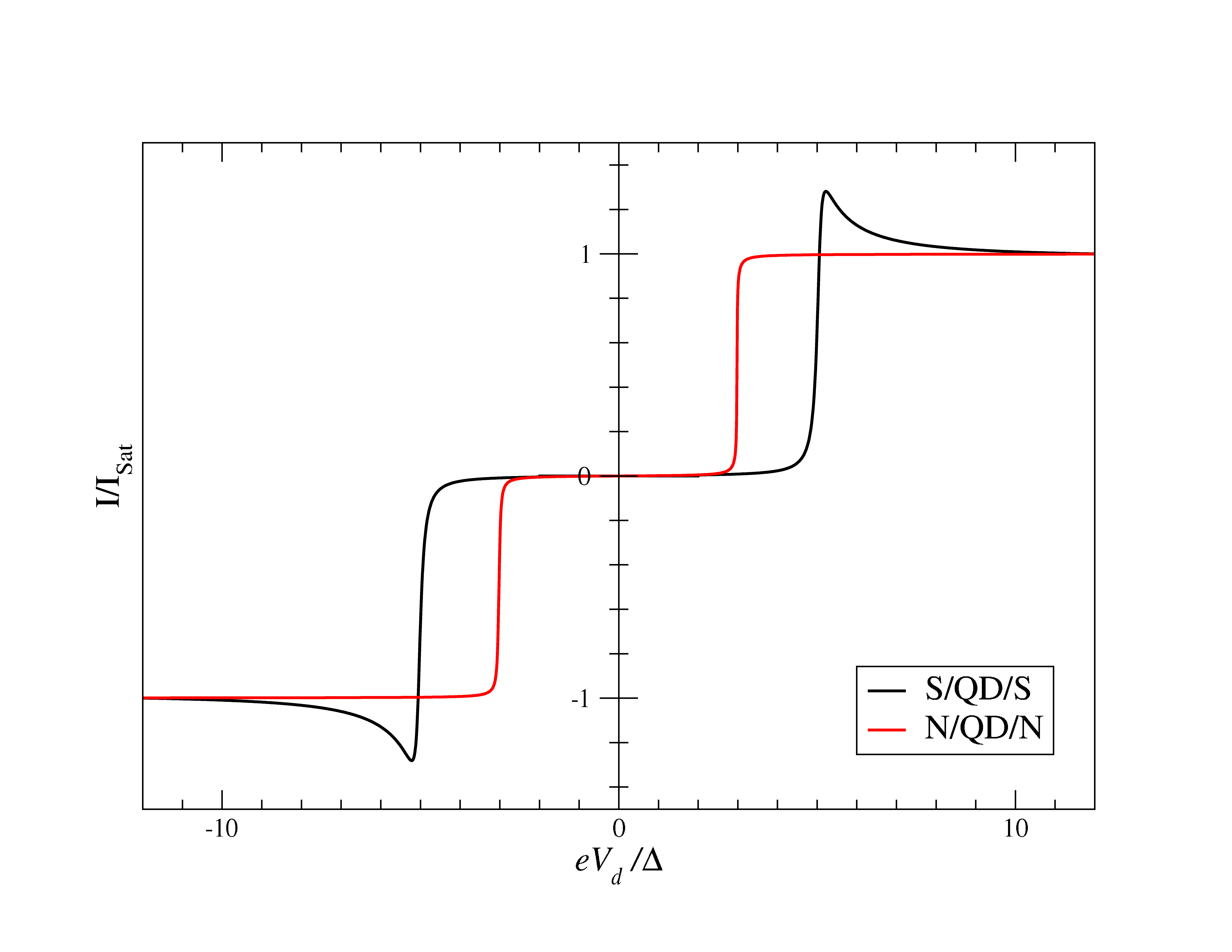}}
\caption{Zero temperature ${\rm I}-V$ characteristics for
superconductor-quantum-dot-superconductor system,
calculated using the self consistent field (SCF)
method, with
${\rm E_{0}} = 1.5~{\rm \Delta}$, $eV_{g} = 0.0~{\rm \Delta}$,
$U_{0}=0.005~{\rm \Delta}$,
$C_{d}/C_{E} = 0.5$,
${{\rm \Gamma}}_{L} =
{{\rm \Gamma}}_{R} = 0.005~{\rm \Delta}$.}
\label{figure3}
\end{figure}

\begin{figure}
\resizebox{1.0\hsize}{!}{\includegraphics*{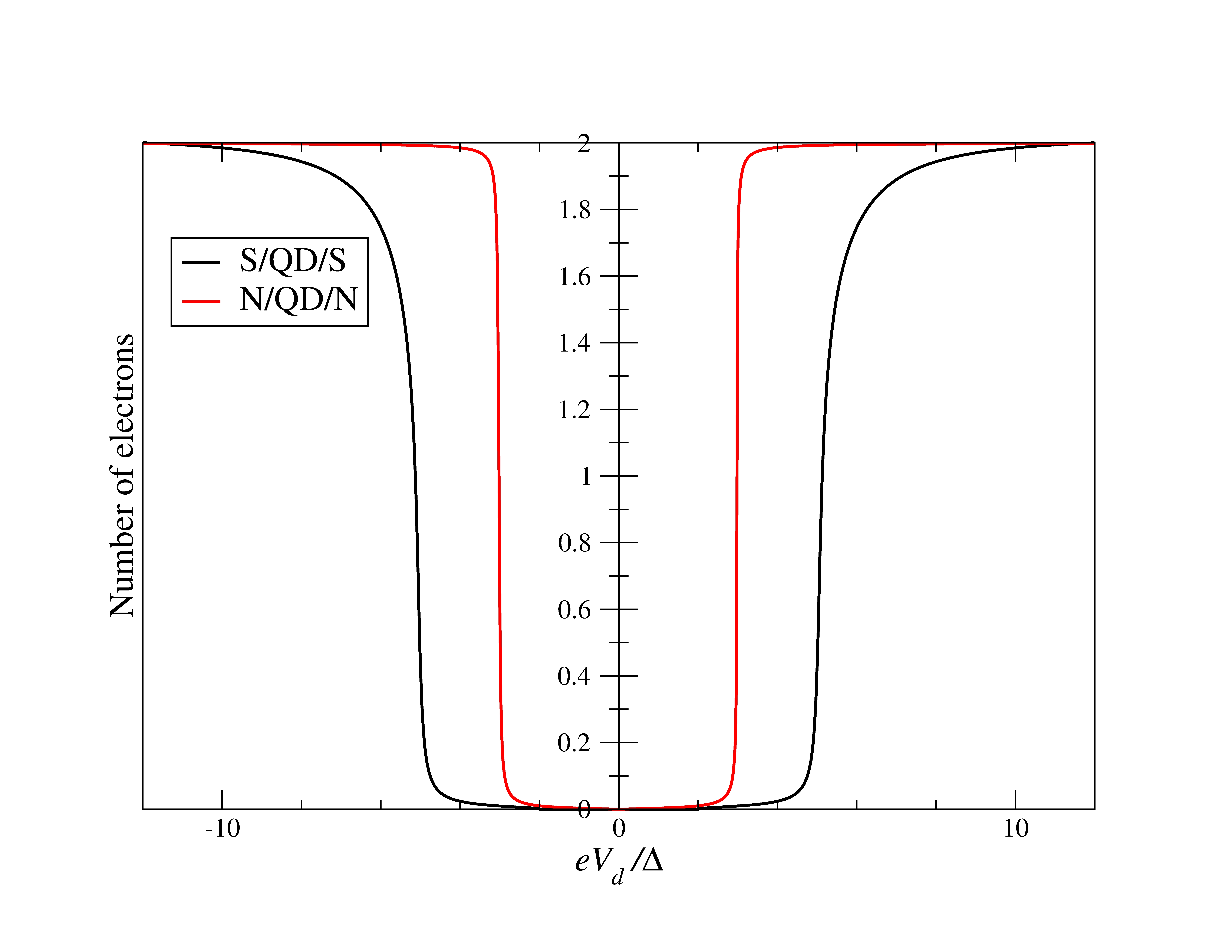}}
\caption{Zero temperature Number of electrons-$eV_{d}/\Delta$ graph for
superconductor-quantum-dot-superconductor system, calculated
using the self consistent
field (SCF) method, with
${\rm E_{0}} = 1.5~{\rm \Delta}$, $eV_{g} = 0.0~{\rm \Delta}$,
$U_{0}=0.005~{\rm \Delta}~$,
$C_{d}/C_{E} = 0.5$,
${{\rm \Gamma}}_{L} =
{{\rm \Gamma}}_{R} = 0.005~{\rm \Delta}$.}
\label{figure4}
\end{figure}
In this regime there are the Multiple Andreev reflections \cite{Andreev1} for voltages such that, $eV<\Delta$ (MAR). Also, there is the possibility for quasi-particle  co-tunneling current for energy levels far from $\mu_{L}$. These cases will be considered in a future work and involves the whole expressions we have derived for the currents (Equations (\ref{equation32}) and (\ref{equation33})) and  eventually more accurate Green-Keldysh functions and the use of master equations \cite{MRoderoI}. This case was studied experimentally in \cite{HTakayanagi}.
For given values of capacitances and source voltage, we iterate Equations (\ref{equation72}) and (\ref{equation74}), in order to find the potential $U$.
Then the single particle current, ${\rm I}_{SP}(V,U)$ is evaluated (Equation (\ref{equation76})).
We put  the charge before biasing $N_{0}=0$, such that Coulomb repulsion with the QD-energy level is absent. Anyhow, in this regime, the effect of the second term in Equation (\ref{equation79}) is negligible. Consequently, the Laplace term $U_{\cal L}$, essentially position the QD degenerate energy level (with respect $\mu_{L}=0$). In Figure \ref{figure3} we show  ${\rm I}-V$
characteristics for gate voltage values $V_{g}=0$ and $K_{B}T\ll\Delta$, whereas Figure \ref{figure4} shows the occupation number $\Delta N$. These curves are symmetric, due to the assumed equality of the coupling capacitances ($C_{d}/C_{E}=0.5$). Otherwise, the ${\rm I}-V$ shifts to right or to the left for $C_{d}/C_{E}=0.5>0.5$ or $< 0.5$ respectively. For this case, we show in Figure \ref{figure11} the spectral density $\rho(\omega)$ for $eV_{d}=-6~\Delta$. Notice that the position of the energy level is essentially $-4.5~\Delta$, i.e., just the sum of $E_{0}+U_{\cal L}$.
Qualitatively, these results are similar of Yeyati et al. \cite{MRoderoI}. Characteristic is the broadening of the BSC singularity.
The effect of bigger values of $\Gamma_{R,L}$ is a more pronounced round off the BCS-type singularity. We discuss this issue below. For large enough  bias the current approaches the normal saturation value ${\rm I_{Sat}}$.

\subsection{Second case: ${\rm \Delta}\simeq U_{\rm 0}\gg {{\rm \Gamma}}_{L,R}$}
\begin{figure}
\resizebox{1.0\hsize}{!}{\includegraphics*{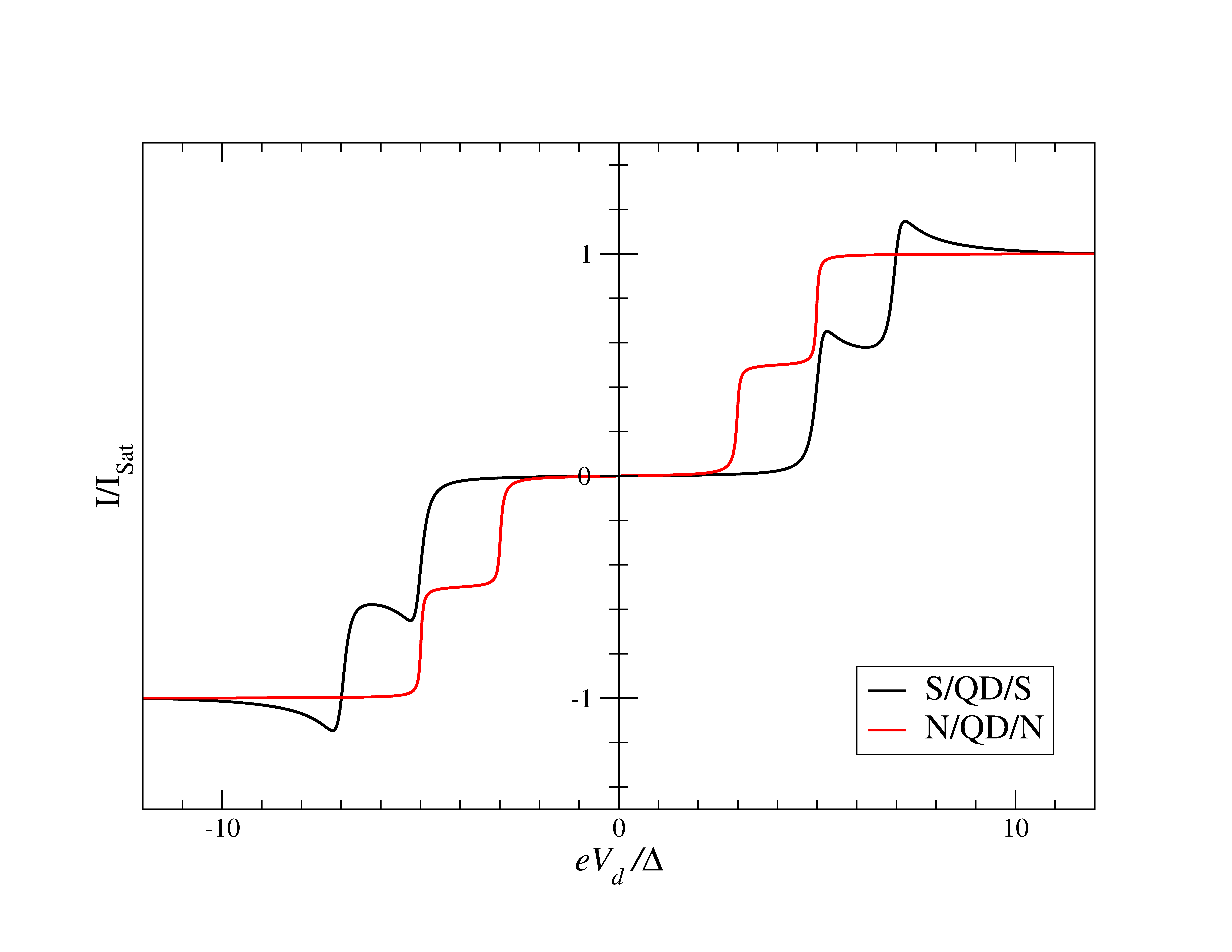}}
\caption{Zero temperature ${\rm I}-V$ characteristics showing the Coulomb blockade for
superconductor-quantum-dot-superconductor system, calculated using the
self consistent field (SCF)
method, with
${\rm E_{0}} = 1.5~{\rm \Delta}$, $eV_{g} = 0.0~{\rm \Delta}$,
$U_{0}=1.0~{\rm \Delta}$,
$C_{d}/C_{E} = 0.5$,
${{\rm \Gamma}}_{L} =
{{\rm \Gamma}}_{R} = 0.01~{\rm \Delta}$.}
\label{figure7}
\end{figure}

\begin{figure}
\resizebox{1.0\hsize}{!}{\includegraphics*{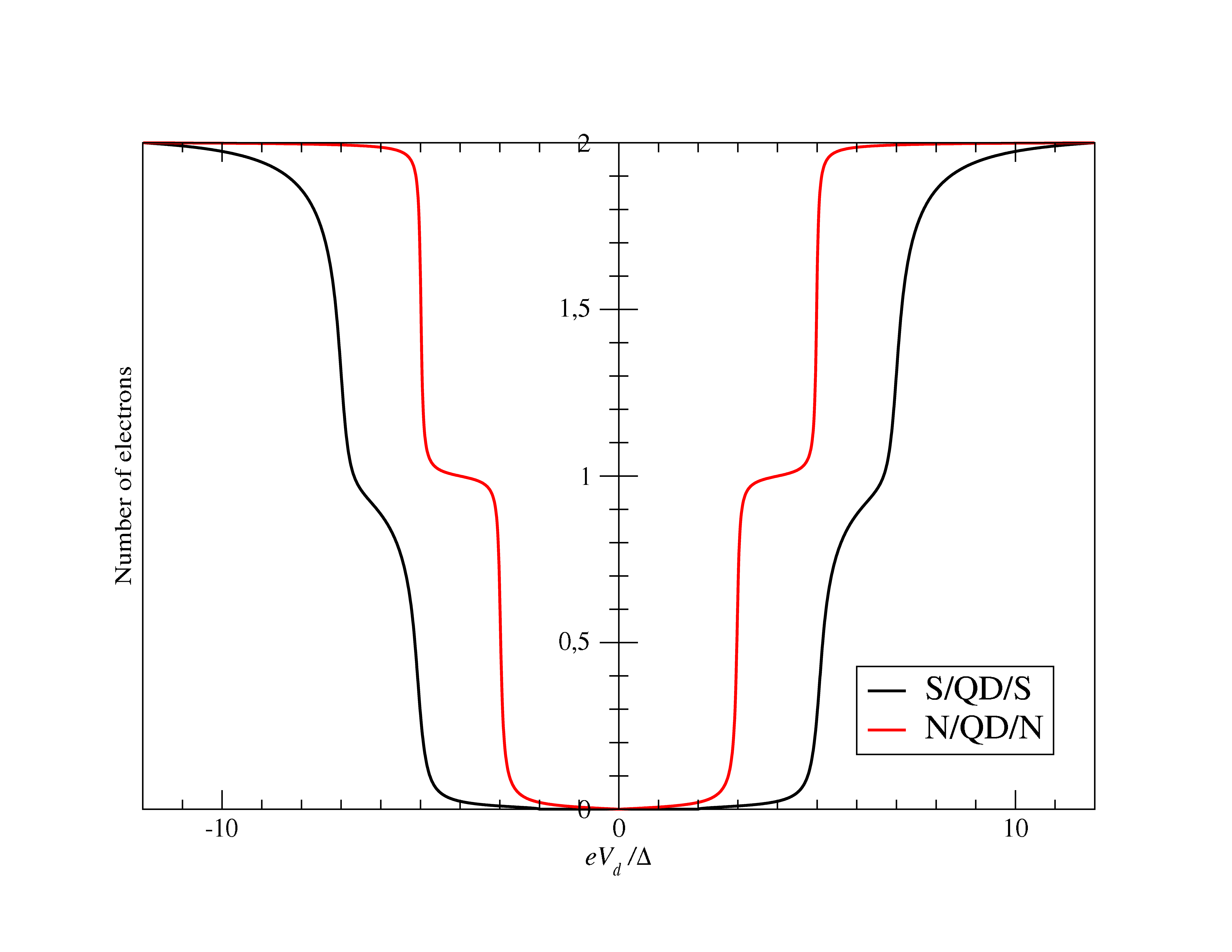}}
\caption{Zero temperature Number of electrons-$eV_{d}/{\rm \Delta}$ for
superconductor-quantum-dot-superconductor system, calculated
using the self consistent field (SCF) method, with
${\rm E_{0}} = 1.5~{\rm \Delta}$, $eV_{g} = 0.0~{\rm \Delta}$,
$U_{0}=1.0~{\rm \Delta}$,
$C_{d}/C_{E} = 0.5$,
${{\rm \Gamma}}_{L} =
{{\rm \Gamma}}_{R} = 0.01~{\rm \Delta}$.}
\label{figure8}
\end{figure}

\begin{figure}
\resizebox{1.0\hsize}{!}{\includegraphics*{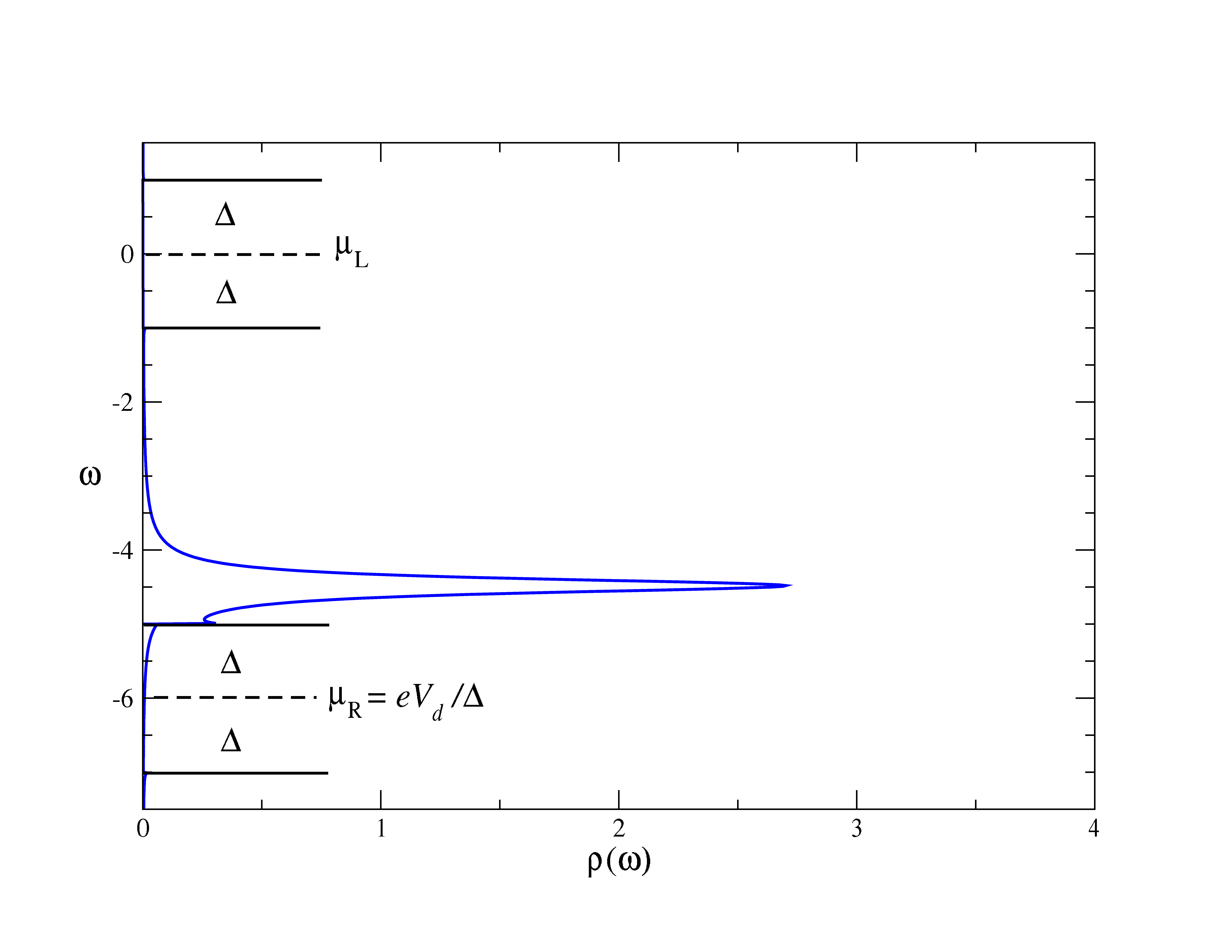}}
\caption{Espectral density of the quantum dot-$\omega$ for
superconductor-quantum-dot-superconductor system, calculated
using the self consistent field (SCF) method, with
${\rm E_{0}} = 1.5~{\rm \Delta}$, $eV_{g} = 0.0~{\rm \Delta}$,
$eV_{d} = 6.0~\Delta$,
$U_{0}=0.01~{\rm \Delta}$,
$C_{d}/C_{E} = 0.5$,
${{\rm \Gamma}}_{L} =
{{\rm \Gamma}}_{R} = 0.05~{\rm \Delta}$.}
\label{figure11}
\end{figure}

\begin{figure}
\resizebox{1.0\hsize}{!}{\includegraphics*{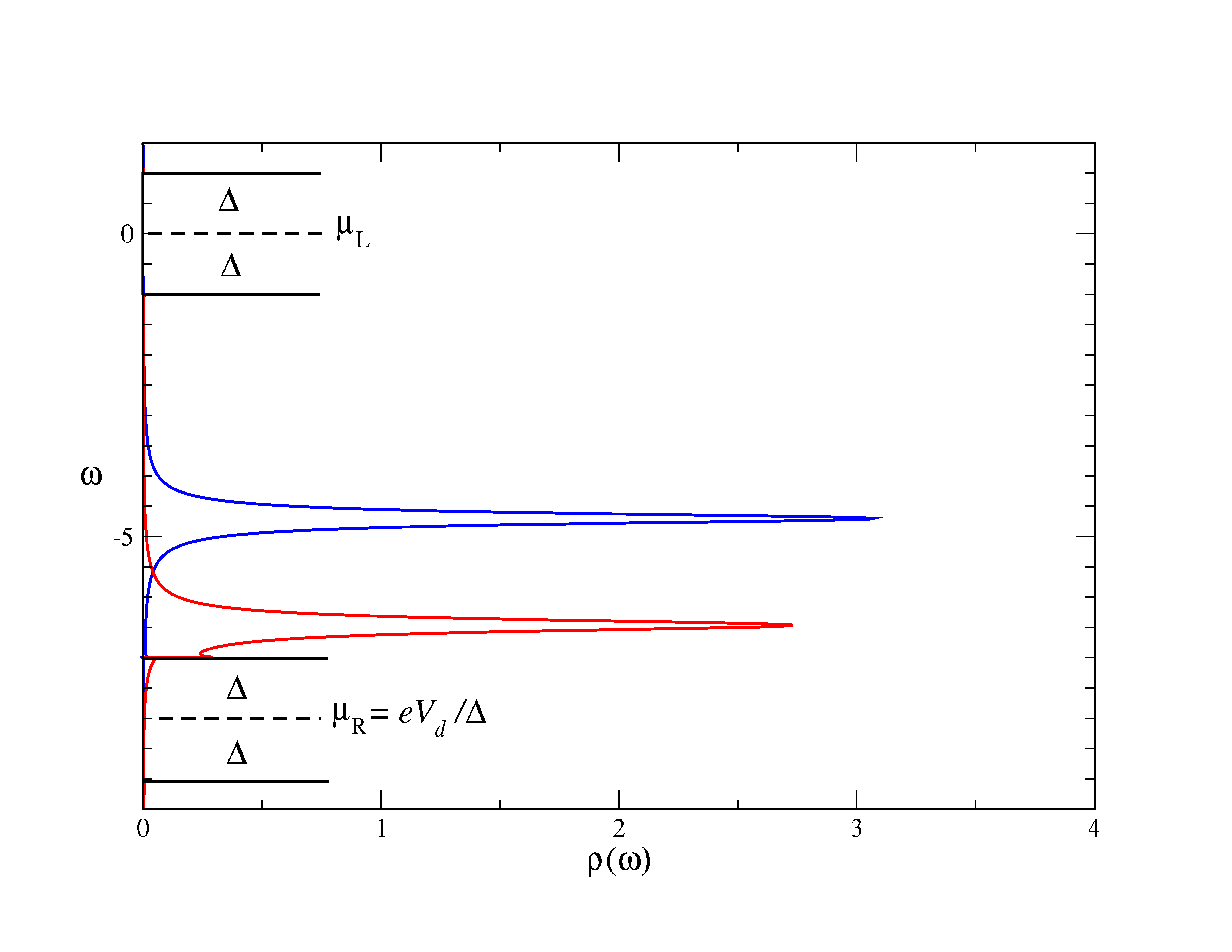}}
\caption{Espectral density of the quantum dot-$\omega$ for
superconductor-quantum-dot-superconductor system, calculated
using the self consistent field (SCF) method, with
${\rm E_{0}} = 1.5~{\rm \Delta}$, $eV_{g} = 0.0~{\rm \Delta}$,
$eV_{d} = 8.0~\Delta$,
$U_{0}=1.0~{\rm \Delta}$,
$C_{d}/C_{E} = 0.5$,
${{\rm \Gamma}}_{L} =
{{\rm \Gamma}}_{R} = 0.20~{\rm \Delta}$.}
\label{figure12}
\end{figure}

In this regime the charging energy acts effectively in lifting the degeneracy of the otherwise single degenerate QD-energy level. For this regime, we use the couple system defined by Equations (\ref{equation77a}) - (\ref{equation80}) and calculate the current according to Equation (\ref{equation81}). This is the unrestricted  SCF method mentioned in the introduction. The transport begins through one level as long as there is in average less than one electron in it. For the given parameters the onset of current is similar to the first case (No interaction with residual charge in the QD is considered). However, when the average occupation exceeds one, the  other degenerate levels floats according to the  resulting values values $U_{\uparrow}$ and $U_{\downarrow}$. This values push down the position of this second level and push up the already occupied  energy level.
In Figures \ref{figure7} - \ref{figure8}, we show ${\rm I}-V$ and the number of electrons. In Figure \ref{figure12} it is shown the spectral density for $eV_{d}=-8\Delta$. In this case, $E_{0}+U_{\cal L}=-5.5\Delta$. The values obtained from the SCF calculation position the energy levels to  $-4.846\Delta$ and to $-6.396\Delta$ (see equation (\ref{equation77b})).


\begin{figure}
\resizebox{1.0\hsize}{!}{\includegraphics*{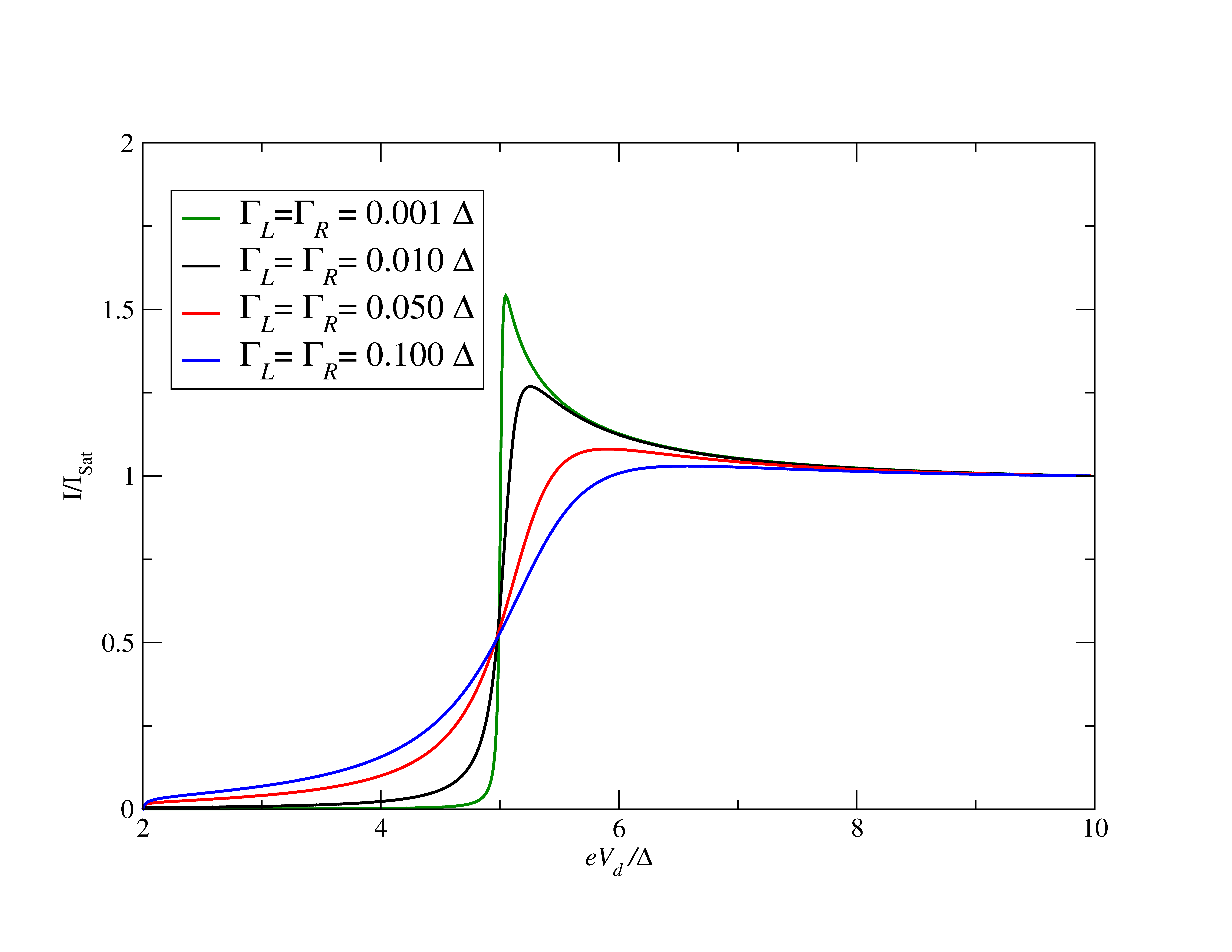}}
\caption{Zero temperature ${\rm I}-V$ characteristics for
 superconductor-quantum-dot-superconductor system
for various values of the coupling ${\rm \Gamma}$, calculated
 using the self consistent field (SCF)
 method, with
 ${\rm E_{0}} = 1.5~{\rm \Delta}$, $eV_{g} = 0.0~{\rm \Delta}$,
 $U_{\rm 0} = 0.05 {\rm \Delta}$,
 $C_{d}/C_{E} = 0.5$,
${{\rm \Gamma}}_{L} = {{\rm \Gamma}}_{R}$.
}
\label{figure13}
\end{figure}
\begin{figure}
\caption{Zero temperature ${\rm I}-V$ characteristics for
 superconductor-quantum-dot-superconductor system
for various values of the coupling ${\rm \Gamma}$, calculated
 using the self consistent field (SCF)
 method, with
 ${\rm E_{0}} = 1.5~{\rm \Delta}$, $eV_{g} = 0.0~{\rm \Delta}$,
 $U_{\rm 0} = 0.05~{\rm \Delta}$,
 $C_{d}/C_{E} = 0.5$,
${\rm \Gamma}_{L} = 4{\rm \Gamma}_{R}$.}
\label{figure14}
\end{figure}

\begin{figure}
\caption{Zero temperature ${\rm I}-V$ characteristics for
 superconductor-quantum-dot-superconductor system
for various values of the coupling ${\rm \Gamma}$, calculated
 using the self consistent field (SCF)
 method, with
 ${\rm E_{0}} = 1.5~{\rm \Delta}$, $eV_{g} = 0.0~{\rm \Delta}$,
 $U_{\rm 0} = 0.05~{\rm \Delta}$,
 $C_{d}/C_{E} = 0.5$,
${\rm \Gamma}_{R} = 4{\rm \Gamma}_{L}$.}
\label{figure15}
\end{figure}

\begin{figure}
\caption{Zero temperature ${\rm I}-V$ characteristics for
 superconductor-quantum-dot-superconductor system
for various values of the coupling ${\rm \Gamma}$, calculated
 using the self consistent field (SCF)
 method, with
 ${\rm E_{0}} = 1.5~{\rm \Delta}$, $eV_{g} = 0.0~{\rm \Delta}$,
 $U_{\rm 0} = 1.0~{\rm \Delta}$,
 $C_{d}/C_{E} = 0.5$,
${\rm \Gamma}_{L} = {\rm \Gamma}_{R}$.}
\label{figure16}
\end{figure}

\begin{figure}
\caption{Zero temperature ${\rm I}-V$ characteristics for
 superconductor-quantum-dot-superconductor system
for various values of the coupling ${\rm \Gamma}$, calculated
 using the self consistent field (SCF)
 method, with
 ${\rm E_{0}} = 1.5~{\rm \Delta}$, $eV_{g} = 0.0~{\rm \Delta}$,
 $U_{\rm 0} = 1.0~{\rm \Delta}$,
 $C_{d}/C_{E} = 0.5$,
${\rm \Gamma}_{L} = 4{\rm \Gamma}_{R}$.}
\label{figure17}
\end{figure}

\begin{figure}
\caption{Zero temperature ${\rm I}-V$ characteristics for
 superconductor-quantum-dot-superconductor system
for various values of the coupling ${\rm \Gamma}$, calculated
 using the self consistent field (SCF)
 method, with
 ${\rm E_{0}} = 1.5~{\rm \Delta}$, $eV_{g} = 0.0~{\rm \Delta}$,
 $U_{\rm 0} = 1.0~{\rm \Delta}$,
 $C_{d}/C_{E} = 0.5$,
${\rm \Gamma}_{R} = 4{\rm \Gamma}_{L}$.}
\label{figure18}
\end{figure}

The experimental work of Ralph et al \cite{RalphBlackTinkham} corresponds to this second case. In their Figure 3, they show the current  for
a single state level peaks ${\rm I}\sim 5$ pA at a bias  $V\sim2.4$ mV. According to our Figures \ref{figure7} - \ref{figure8}, for this
sample at $30$ mK,
there should be another peak at $V\sim4.4$ mV with a current value ${\rm I}_{SP}\simeq10$ pA.  The charging energy for this sample is
$U_{0}\simeq2.0$ mV. However, in their samples with radios $\sim 2.5$ nm or greater, the level spacing of the energy levels is such, that
$\Delta E < E_{c}$, therefore another current signal may occurs before and a quantitative  proper description would be a multilevel QD-
model. One notice, however, the strong fluctuations in the spacing in Figure 2 \cite{RalphBlackTinkham}, indicating complex charging many
levels phenomena. Notice that the theoretical explanation of Yeyati et al. \cite{MRoderoI,MRoderoII}, does not contain our prediction.

\subsection{Influence of Coupling Constants.}
The influence of the coupling constant is to broaden the otherwise sharp energy QD level. However, the broadening is not equally strong, and
depends on the relative values of $\Gamma_{R}/\Gamma_{L}$. Ralph et al. \cite{RalphBlackTinkham} determined for the sample in his Figure 3
$\Gamma_{R}/\Gamma_{L}$ $\gg1$. Figure \ref{figure14} shows the ${\rm I}-V$ characteristics for the restricted case when both coupling
constants are equal. For larger values of the  coupling constants the broadening is stronger. If they are dissimilar in value, the
broadening is stronger when $\Gamma_{R}>\Gamma_{L}$. This effect is shown in Figures \ref{figure13} - \ref{figure18}. This effect is due to
the stronger involving of the BSC-DOS singularity of the left lead in the integral expression for the current (Equation (\ref{equation66}))
and the particular choice of the zero  bias voltage (see Figure \ref{figure2}).

\section{SUMMARY and PERSPECTIVES}\label{conclusions}
We have studied the single particle current through a quantum dot coupled with two superconductor leads via a coupled  Poisson Non-
equilibrium Green Function (PNEGF) formalism. In a systematic and self contained way, we derived the expressions for the current in
full generality. In this work we focused  only in the weak coupling regime where single particle current is dominant one. The QD is
a single degenerate energy level systems modeled via a capacitive circuit. The influence of the Potential on the QD, on the ${\rm I}-V$
characteristic is calculated for  relevance values of the coupling and capacitances and the implication for experiments is discussed. This
was done in the weak coupling regime and for  ${\rm \Delta} \gg {{\rm \Gamma}}_{L,R}\simeq U_{\rm 0}$. A second case when ${\rm
\Delta}\simeq U_{\rm 0}\gg {{\rm \Gamma}}_{L,R}$ also in the weak coupling regime was analyzed.
Admittedly, our model of a Hybrid system S/QD/S possess potentially  physical extensions. One important missed point is dephasing. This
physical effect due to scattering of transport electrons can be incorporated in the self energy phenomenologically
(\cite{SDattaDephasing1,SDattaDephasing2}), or in a stochastic fashion (\cite{RRangelEMedina}). Another point is to consider a QD with many
energy levels and within the self consistent scheme, to consider the strong and intermediate regimes and many body correlations due to
different  kinds of electron-electron interaction. Here we have to notice that is it not just to scale the level spacing by the charging
energy \cite{WhanOrlando}. It is a genuine many body problem. But the most important missed point was correlations. As pointed out by Datta
(chapter III \cite{SDattaBook2005}), there has been much effort in order to find a suitable SCF that considers correlations.  For example,
to modify Equation (\ref{equation72}) to consider occupancies probabilities.  As discussed in the introduction, Kang \cite{KKang} and Meir
et al. \cite{MeirI} finds a solution for the QD Green function that contains this type of correlation.  In other words, one could go to
scheme where a more accurate Green function for the QD is used together with a multi-electron picture with associated Master equation. These
would mean to use the Anderson model with a Coulomb interaction $U$ that is obtained from a SCF. We want to check if this point of view is
correct.
Work in this direction is in progress.

\newpage
\section{Appendix A}

\begin{figure}
\caption{The contour $ {\rm C_{K}}={\rm C_{K_{-}}}\cup {\rm C_{K_{+}}}$ runs on the real axis, but for clarity its two branches ${\rm
C_{K_{-}}}$ and ${\rm C_{K_{+}}}$ are shown slightly away from the real axis. The contour $ {\rm C_{K}}$ runs from $ t_0$ and return to $
t_0$.}
\label{figure19}
\end{figure}

In general, any Green-Keldysh function  of two operators $A(t)$ and $B(t)$ function is given by:
\begin{equation}
{\rm G}_{A,B}{\left(t, t'\right)}=-{\rm i}\langle {\rm T_{{K}}}A\left(t\right)B\left(t'\right)\rangle
\label{GreenFunctionAB}
\end{equation}
where the operator ${\rm T_{K}}$ acts on the Keldysh contour shown in Figure \ref{figure15}.
A Heaviside function on the Keldysh contour is given by:

\begin{equation}
{\rm \Theta}\left(t,t'\right)
\equiv  \left\{ \begin{array}{cl}
         {\rm \Theta}\left(t-t'\right), & \mbox{$\quad t\in {\rm C_{K_{-}}},\,\,t'\in {\rm C_{K_{-}}}$};\\
        &\\
        0,& \mbox{$\quad t\in {\rm C_{K_{-}}},\,\,t'\in {\rm C_{K_{+}}}$};\\
        &\\
        1,& \mbox{$\quad t\in {\rm C_{K_{+}}},\,\,t'\in {\rm C_{K_{-}}}$};\\
        &\\
        {\rm \Theta}\left(t'-t\right),& \mbox{$\quad t\in {\rm C_{K_{+}}},\,\,t'\in {\rm C_{K_{+}}}$}.
        \end{array} \right.
        \label{stepfunction}
\end{equation}
whereas  be derivative of the Heaviside function on the Keldysh contour is given by:
\begin{equation}
 \delta\left(t,t'\right)
\equiv
\frac{\partial{\rm \Theta}\left(t,t'\right)}{\partial t}
=
-\frac{\partial{\rm \Theta}\left(t,t'\right)}{\partial t'}
=
\left\{ \begin{array}{cl}
         \delta\left(t-t'\right), & \mbox{$\quad t\in {\rm C_{K_{-}}},\,\,t'\in {\rm C_{K_{-}}}$};\\
        &\\
        0,& \mbox{$\quad t\in {\rm C_{K_{-}}},\,\,t'\in {\rm C_{K_{+}}}$};\\
        &\\
        0,& \mbox{$\quad t\in {\rm C_{K_{+}}},\,\,t'\in {\rm C_{K_{-}}}$};\\
        &\\
        -\delta\left(t-t'\right),& \mbox{$\quad t\in {\rm C_{K_{+}}},\,\,t'\in {\rm C_{K_{+}}}$}.
        \end{array} \right.
        \label{detafunction}
\end{equation}
In general a function ${\rm F}\left(t,t\right)$ defined on the Keldsyh contour is given by:

\begin{equation}
{\rm F}\left(t,t'\right)
\equiv
{\rm \Theta}\left(t,t'\right){\rm F}^{>}\left(t,t'\right)
+
{\rm \Theta}\left(t',t\right){\rm F}^{<}\left(t,t'\right)
\end{equation}
where ${\rm F}^{>}\left(t,t'\right)$ is the so called greater component (greater Keldysh-Green function) and ${\rm F}^{<}\left(t,t'\right)$
is the lesser component (lesser Keldysh-Green function) of ${\rm F}\left(t,t'\right)$.

Directly calculations can be carried out, in this way by deriving by $t$ or $t'$  and using Heisenberg equation
of motion for the time evolution of the operators, $A(t)=\exp({iHt})A\exp({-iHt})$,
\begin{equation}
\frac{\partial {\rm A}\left(t\right)}{\partial t}
=
-{\rm i}\left\lbrack A\left(t\right),H\right\rbrack,
\end{equation}
ones obtains:
\begin{equation}
{\rm i}\frac{\partial {\rm G}\left(t,t'\right)}{\partial t}
=
\delta\left(t,t'\right)\left\langle\left\lbrack A\left(t\right),B\left(t\right)\right\rbrack_{\mp}\right\rangle
-
{\rm i}\left\langle{\rm T_K}\left\lbrack A\left(t\right),H\right\rbrack B\left(t'\right)\right\rangle,
\end{equation}
\begin{equation}
-{\rm i}\frac{\partial {\rm G}\left(t,t'\right)}{\partial t}
=
\delta\left(t,t'\right)\left\langle\left\lbrack A\left(t\right),B\left(t\right)\right\rbrack_{\mp}\right\rangle
+
{\rm i}\left\langle{\rm T_K} A\left(t\right)\left\lbrack B\left(t'\right),H\right\rbrack \right\rangle.
\end{equation}
If there is not an applied potential, i.e., if $V_{s}=V_{d}=0$, (see Figure \ref{figure2}), the whole system is at equilibrium, one can use
the usual commutator
Green Functions, or equivalently, the Matsubara or Temperature-Green Function to quantify correlations functions. In that case, Equation
(\ref{GreenFunctionAB}) depends only on the time difference $(t-t')$. However, for times $t>t_{0}$, once the potential  difference has been
applied, the  simple dependence on the time differences not longer holds which signalizes a  non equilibrium situation. In that case the
Keldysh method applies.

\section{Appendix B}
We encounter two cases (Equations (\ref{equation28}) and (\ref{equation44})), where
the general strategy to find an integral expression for the Keldyhs-Green Functions is the following:
1) one first considers the equation of motion for  for the Keldysh-Green function of each systems (two leads and the QD) (Equation
(\ref{equation27}) and Equation (\ref{equation40})).
2) The resulting equation of motion in each case has a delta function as inhomogeneity.
3) The systems are connected in at $t_{0}$, we have as a result two coupled equations of motion (Equations (\ref{equation20}) and
(\ref{equation43})).
4) These equations are converted into  an equivalent integral equation on the Keldysh contour.
5) Straightforwardly derivation od the integral equations results in the differential equation of motion.

\section{Appendix C}
In this Appendix we explain how we evaluate the some important convolutions used to calculate lesser Keldysh-Green functions
\cite{DCLangrethI,DCLangrethII}.
In general a function given as a convolution on the Keldysh contour poses the definition given in Appendix A.
One encounter situations where a Keldysh-Green function is given by a convolution of two other functions:

$$
{\rm P}\left(t,t'\right)
=
\int_{{\rm C_{K}}}{\rm d''}{\rm F}\left(t,t''\right){\rm G}\left(t'',t'\right).
$$
However, for evaluation of quantities like for example the current, one needs
$$
{\rm P}^{<}\left(t,t'\right)
=
\int_{{\rm C_{K}}}{\rm d''}{\rm F}\left(t,t''\right){\rm G}\left(t'',t'\right)\arrowvert_{t<_{\rm C_{K}}t'}.
$$
One sees that, the relative position of $t$ and $t'$ divides the contour in three regions of integration:
1) $t''<_{\rm C_{K}}t$,  2) $t<_{\rm C_{K}}t''<t'$ and 3) $t''>_{\rm C_{K}}t'$.
This traduces into the following integrals:
\begin{eqnarray}
&&{\rm P}^{<}\left(t,t'\right)
=
\int^{t}_{t_{0}}{\rm d}t''{\rm F}^{>}\left(t,t''\right){\rm G}^{<}\left(t'',t'\right)|_{t''<_{\rm C_{K}}t}+\nonumber\\
&&\int^{t'}_{t}{\rm d}t''{\rm F}^{<}\left(t,t''\right){\rm G}^{<}\left(t'',t'\right)|_{t<_{\rm C_{K}}t''<t'}
+
\int^{t_{0}}_{t'}{\rm d}t''{\rm F}^{<}\left(t,t''\right){\rm G}^{>}\left(t'',t'\right)|_{t''>_{\rm C_{K}}t'}.
\label{appc1}
\end{eqnarray}

With book-keeping manipulations of the second integral, one obtains:
\begin{equation}
{\rm P}^{<}\left(t,t'\right)
=
\int^{t_{1}}_{t_{0}}{\rm d}t''\left\lbrack{\rm F}^{\left(\rm r\right)}\left(t,t''\right){\rm G}^{<}\left(t'',t'\right)+
{\rm F}^{<}\left(t,t''\right){\rm G}^{\left({\rm a}\right)}\left(t'',t'\right)\right\rbrack\nonumber,
\label{appc2}
\end{equation}
where the retarded Keldysh-Green function ${\rm F}^{\left(\rm r\right)}\left(t,t''\right)$
and the advance Keldysh-Green function ${\rm G}^{\left({\rm a}\right)}\left(t,t''\right)$ are given by:
\begin{eqnarray}
{\rm F}^{\left(\rm r\right)}\left(t,t''\right)
&=&
{\rm \Theta}\left(t-t''\right)[{\rm F}^{>}\left(t,t''\right)-
{\rm F}^{<}\left(t,t''\right)],\nonumber\\
\label{appc3}
{\rm G}^{\left({\rm a}\right)}\left(t'',t'\right)
&=&
-{\rm \Theta}\left(t'-t''\right)[{\rm G}^{>}\left(t'',t'\right)-
{\rm G}^{<}\left(t'',t'\right)].
\label{appc4}
\end{eqnarray}

In the say way of reasoning one obtains:
\begin{equation}
{\rm P}^{<\left\{>\right\}}\left(t,t'\right)
=
\int^{\infty}_{-\infty}{\rm d}t''\left[{\rm F}^{\left(\rm r\right)}\left(t,t''\right){\rm G}^{<\left\{>\right\}}\left(t'',t'\right)+{\rm
F}^{<\left\{>\right\}}\left(t,t''\right){\rm G}^{\left(\rm a\right)}\left(t'',t'\right)\right].
\label{appc5}
\end{equation}
With the definitions of a retarded/advanced Kedysh-Green function one easily obtains for ${\rm P}^{\left(\rm r\right)}\left(t,t'\right)$:
\begin{eqnarray*}
{\rm P}^{\left(\rm r\right)}\left(t,t'\right)
=
&{\rm \Theta}&\left(t-t'\right)\int^{\infty}_{-\infty}{\rm d}t'' {\rm F}^{\left(\rm r\right)}\left(t,t''\right)\left[{\rm G}^{>}\left(t''-
t'\right)-{\rm G}^{<}\left(t'',t'\right)\right] \\
&+&\\
&{\rm \Theta}&\left(t'-t\right)\int^{\infty}_{-\infty}{\rm d}t''\left[{\rm F}^{>}\left(t-t''\right)-{\rm F}^{<}\left(t,t''\right)\right]
{\rm G}^{\left(\rm r\right)}\left(t'',t'\right),
\end{eqnarray*}
and therefore:
\begin{equation}
{\rm P}^{\left(\rm r\right)}\left(t,t'\right)
=
\int^{\infty}_{-\infty}{\rm d}t''{\rm F}^{\left(\rm r\right)}\left(t,t''\right){\rm G}^{\left(\rm r\right)}\left(t'',t'\right).
\label{appc6}
\end{equation}

Similarly for ${\rm P}^{\left(\rm a\right)}\left(t,t'\right)$ one obtains:
\begin{equation}
{\rm P}^{\left(\rm a\right)}\left(t,t'\right)
=
\int^{\infty}_{-\infty}{\rm d}t''{\rm F}^{\left(\rm a\right)}\left(t,t''\right){\rm G}^{\left(\rm a\right)}\left(t'',t'\right).
\label{appc7}
\end{equation}

\section{Appendix D}
Below we proceed to evaluate the unperturbed Green's functions ${\rm g}_{\eta\vec{k}\sigma}$ and ${\rm f}_{\eta\vec{k}\sigma}$. To achieve
this, it is necessary to introduce the chemical potential shift in each superconductor,
$$
{\mathcal H}_{\eta}
=
H_{\eta}-\mu_{\eta}N_{\eta},
$$
so that
\begin{eqnarray*}
a_{\eta\vec{k}\sigma}\left(t\right)
\equiv
{\rm e}^{{\rm i}H_{\eta}t}a_{\eta\vec{k}\sigma}{\rm e}^{-{\rm i}H_{\eta}t}
&=&
{\rm e}^{{\rm i}{\mathcal H}_{\eta}t}\left({\rm e}^{{\rm i}\mu_{\eta}t}a_{\eta\vec{k}\sigma}{\rm e}^{-{\rm i}\mu_{\eta}t}\right){\rm e}^{-{\rm i}{\mathcal H}_{\eta}t}\\
&=&{\rm e}^{{-\rm i}\mu_{\eta}t}\left({\rm e}^{{\rm i}{\mathcal H}_{\eta}t}a_{\eta\vec{k}\sigma}{\rm e}^{-{\rm i}{\mathcal H}_{\eta}t}\right) \rightarrow
{\rm e}^{{-\rm i}\mu_{\eta}t}a_{\eta\vec{k}\sigma}\left(t\right),
\end{eqnarray*}
due to
\begin{eqnarray*}
{\rm e}^{{\rm i}{\mathcal H}_{\eta}t}a_{\eta\vec{k}\sigma}{\rm e}^{-{\rm i}{\mathcal H}_{\eta}t}
&=&
a_{\eta\vec{k}\sigma}{\rm e}^{-{\rm i}{\rm E}_{\eta\vec{k}}t},\\
{\rm e}^{{\rm i}{N}_{\eta}t}a_{\eta\vec{k}\sigma}{\rm e}^{-{\rm i}{N}_{\eta}t}
&=&
a_{\eta\vec{k}\sigma}{\rm e}^{-{\rm i}{\mu}_{\eta\vec{k}}t},
\end{eqnarray*}
because $\left\lbrack N_{\eta},H_{\eta}\right\rbrack=0$.
Consequently, the first unperturbed retarded Green function ${\rm g}^{({\rm r})}_{\eta\vec{k}\sigma}\left(t,t'\right)$ is given by
\begin{eqnarray*}
{\rm g}^{({\rm r})}_{\eta\vec{k}\sigma}\left(t,t'\right)
&=&
-{\rm i}{\rm \Theta}\left(t-t'\right)\left\langle\left\{{\rm e}^{-{\rm i}\mu_{\eta}t}a_{\eta\vec{k}\sigma}\left(t\right), {\rm e}^{{\rm
i}\mu_{\eta}t'}a{\dagger}_{\eta\vec{k}\sigma}\left(t'\right)\right\}\right\rangle
\\
&=&
-{\rm i}{\rm \Theta}\left(t-t'\right){\rm e}^{-{\rm i}\mu_{\eta}\left(t-t'\right)}
\times\\
&&\left\langle\left\{u_{\eta\vec{k}}\gamma_{\eta\vec{k}\sigma}\left(t\right)+\sigma
v_{\eta\vec{k}}\gamma{\dagger}_{\eta\vec{k}\sigma}\left(t\right), u_{\eta\vec{k}}\gamma{\dagger}_{\eta\vec{k}\sigma}\left(t'\right) + \sigma
v_{\eta\vec{k}}\gamma_{\eta-\vec{k}-\sigma}\left(t'\right)\right\}\right\rangle
\\
&=&
-{\rm i}{\rm \Theta}\left(t-t'\right){\rm e}^{-{\rm i}\mu_{\eta}\left(t-t'\right)}\left[u^{2}_{\eta\vec{k}}{\rm e}^{-{\rm i}{\rm
E}_{\eta\vec{k}}\left(t-t'\right)}+
v^{2}_{\eta\vec{k}}{\rm e}^{{\rm i}{\rm E}_{\eta\vec{k}}\left(t-t'\right)}\right],
\\
&=&
-{\rm i}{\rm \Theta}\left(t-t'\right)\left[u_{\eta\vec{k}}^{2}{{\rm e}^{-{\rm i}\left(E_{\eta\vec{k}}+\mu_{\eta}\right)\left(t-
t'\right)}}+v_{\eta\vec{k}}^{2}{{\rm e}^{{\rm i}\left({\rm E}_{\eta\vec{k}}-\mu_{\eta}\right)\left(t-t'\right)}} \right].
\label{appD1}
\end{eqnarray*}
where the fermion operators $\gamma^{\dagger}_{\eta\vec{k}\sigma}$, $\gamma_{\eta\vec{k}\sigma}$ create and annihilate the ``Bogoliubov
quasi-particles'' and $\sigma=\left\{\begin{array}{cc}\uparrow&=1\\ \downarrow&=-1
\end{array}\right.$ the spin index. They will be linear combinations of the creation and annihilation operators of the real electrons:
\begin{eqnarray*}
a_{\eta\vec{k}\sigma}\left(t\right)
&=&
u_{\eta\vec{k}}\gamma_{\eta\vec{k}\sigma}\left(t\right)+\sigma v_{\eta\vec{k}}\gamma{\dagger}_{\eta\vec{k}\sigma}\left(t\right),
\\
a{\dagger}_{\eta\vec{k}\sigma}\left(t'\right)
&=&
u_{\eta\vec{k}}\gamma{\dagger}_{\eta\vec{k}\sigma}\left(t'\right) + \sigma v_{\eta\vec{k}}\gamma_{\eta-\vec{k}-\sigma}\left(t'\right).
\label{appD2}
\end{eqnarray*}

Applying the Fourier transformations to ${\rm g}^{({\rm r})}_{\eta\vec{k}\sigma}\left(t,t'\right)$
$$
{\rm g}^{({\rm r})}_{\eta\vec{k}}\left(t,t'\right)
=
\int^{\infty}_{-\infty}\frac{{\rm d}\omega}{2\pi}{\rm e}^{-{\rm i}\omega\left(t-t'\right)}{\rm g}^{({\rm
r})}_{\eta\vec{k}}\left(\omega\right),
$$
with
$$
{\rm g}^{({\rm r})}_{\eta\vec{k}}\left(\omega\right)
=
\frac{u^{2}_{\eta\vec{k}}}{\omega-{\rm E}_{\eta\vec{k}}-\mu_{\eta}+{\rm i}0^{+}}
+
\frac{v^{2}_{\eta\vec{k}}}{\omega+{\rm E}_{\eta\vec{k}}-\mu_{\eta}+{\rm i}0^{+}}.
$$
therefore we have
\begin{eqnarray*}
V^{2}_{\eta}\sum_{\vec{k}}{\rm g}^{\left(\rm r\right)}_{\eta\vec{k}}\left(\omega\right)
&=&
\frac{1}{2}V^{2}_{\eta}\sum_{\vec{k}}\left(\frac{1}{\omega-{\rm E}_{\eta\vec{k}}-\mu_{\eta}+{\rm i}0^{+}}+\frac{1}{\omega+{\rm
E}_{\eta\vec{k}}-\mu_{\eta}+{\rm i}0^{+}}\right)
\\
&=&
V^{2}_{\eta}{\rm P}\sum_{\vec{k}}\frac{\omega-\mu_{\eta}}{\left(\omega-\mu_{\eta}\right)^{2}-{\rm E}^{2}_{\eta\vec{k}}}\\
&-&
\frac{1}{2}{\rm i}\pi V^{2}_{\eta}\sum_{\vec{k}}\left[\delta\left(\omega-\mu_{\eta}-{\rm
E}_{\eta\vec{k}}\right)+\delta\left(\omega-\mu_{\eta}+{\rm E}_{\eta\vec{k}}\right)\right]
\\
&=&
-N_{\eta}\left(0\right)V^{2}_{\eta}{\rm P}\int^{\infty}_{-\infty}{\rm d}\xi\frac{\omega-\mu_{\eta}}{\xi^{2}+{\rm
\Delta}^{2}-\left(\omega-\mu\right)^{2}}
-
\frac{1}{2}{\rm i}\pi V^{2}_{\eta}\sum_{\vec{k}}\delta\left(\left\arrowvert\omega-\mu_{\eta}\right\arrowvert-{\rm E}_{\eta\vec{k}}\right).
\label{appD3}
\end{eqnarray*}

Finally

\fbox{\parbox{12cm}{
$$
{\rm \Sigma}^{\left(\rm r\right)}\left(\omega\right)
\equiv
V^{2}_{\eta}\sum_{\vec{k}}{\rm g}^{\left(\rm r\right)}_{\eta\vec{k}}\left(\omega\right)
=
-{\rm \Gamma}_{\eta}\left\lbrack\frac{\omega-\mu_{\eta}}{{\rm \Delta}_{\eta}}\zeta\left({\rm \Delta}_{\eta},\omega-\mu_{\eta}\right)
+
{\rm i}\zeta\left(\omega-\mu_{\eta},{\rm \Delta}_{\eta}\right)\right\rbrack.
$$
}}
\vspace{1cm}

where
$$
{\rm \Gamma}_{\eta}
=
\pi N_{\eta}\left(\omega - \mu_{\eta}\right)V_{\eta}^{2}
\approx
\pi N_{\eta}\left(0\right)V_{\eta}^{2},
\qquad\zeta\left(\omega,\omega'\right)
\equiv
{\rm \Theta}\left(\left\arrowvert\omega|-|\omega'\right\arrowvert\right)\frac{\left\arrowvert\omega\right\arrowvert}
{\sqrt{\omega^{2}-\omega'^{2}}}.
$$
${{\rm \Gamma}}_{\eta}
=
\pi N_{\eta}\left(\omega - \mu_{\eta}\right)V_{\eta}^{2}
\approx
\pi N_{\eta}\left(0\right)V_{\eta}^{2}$.
are the coupling constants between the leads and the
quantum dot in the wide band limit (WBL). $N_{\eta}\left(0\right)$ is the density of states
at the $\eta$ Fermi level and $\mathrm{f}\left(\omega\right)$
is the Fermi-Dirac distribution function.

The  WBL means that the width of the electronic energy bands of the leads are the largest energy.
The density of states in the contacts vary on a scale of Fermi energy.
These scales are of order~$1-10$~eV  $\left(\sim 10^{4}-10^{5}\ {\rm k}\right)$ which are much
larger than the energies involved in the quantum dot~$\sim~{\rm meV} \sim 10~{\rm K}$.
Furthermore, ${\rm \Gamma}_{\eta}\left(\omega - \mu_{\eta}\right)$~varies slowly with~$\omega - \mu_{\eta}$
and the prefactor~${\mathcal D}_{\eta}\left(\omega - \mu_{\eta}\right)$ varies in the range of wide band
and changes~$\omega - \mu_{\eta}$ on the average of~$\arrowvert{V_{\eta\vec{k}}}\arrowvert^{2}$ occur
on the order of~meV. Therefore, we ignore the dependence of~$\omega - \mu_{\eta}$
in~${\rm \Gamma}_{\eta}\left(\omega - \mu_{\eta}\right)$. The WBL establish that an electron in the dot decays
in an continuum of states of the leads and is sufficient condition for the existence of a stationary state,
as has been shown rigourously in \cite{CAPillet}.

\section{Appendix E}
\begin{eqnarray*}
{\rm g}^{<}_{\eta\vec{k}\sigma}\left(t,t' \right)
&=&
{\rm i}\left\langle\left\{{\rm e}^{{\rm i}\mu_{\eta}t'}a{\dagger}_{\eta\vec{k}\sigma}\left(t'\right), {\rm e}^{-{\rm
i}\mu_{\eta}t}a_{\eta\vec{k}\sigma}\left(t\right)\right\}\right\rangle
\\
&=&
{\rm i}{\rm e}^{{\rm i}\mu_{\eta}\left(t-t'\right)}
\left\langle\left[u_{\eta\vec{k}}\gamma{\dagger}_{\eta\vec{k}\sigma}\left(t'\right)+\sigma v_{\eta
\vec{k}}\gamma_{\eta-\vec{k}-\sigma}\left(t'\right)\right]\right.\times
\\
&&\left.\left[ u_{\eta\vec{k}}\gamma_{\eta\vec{k}\sigma}\left(t\right) + \sigma v_{\eta
\vec{k}}\gamma{\dagger}_{\eta-\vec{k}-\sigma}\left(t\right)\right]\right\rangle
\\
&=&
{\rm i}\left[u_{\eta\vec{k}}^{2}{{\rm e}^{-{\rm i}\left({\rm E}_{\eta\vec{k}}+\mu_{\eta}\right)\left(t-t'\right)}}{\rm f}\left({\rm
E}_{\eta\vec{k}}\right)+v_{\eta\vec{k}}^{2}{\rm e}^{{\rm i}\left({\rm E}_{\eta\vec{k}}-\mu_{\eta}\right)\left(t-t'\right)}{\rm f}\left(-{\rm
E}_{\eta\vec{k}}\right) \right] .
\end{eqnarray*}
Applying the Fourier transformations
$$
{\rm g}^{<}_{\eta\vec{k}\sigma}\left(t,t'\right)
=
\int^{\infty}_{-\infty}\frac{{\rm d}\omega}{2\pi}{\rm e}^{-{\rm i}\omega\left(t-t'\right)}{\rm g}^{<}_{\eta\vec{k}}\left(\omega\right),
$$
with
$$
{\rm g}^{<}_{\eta\vec{k}\sigma}\left(\omega\right)
=
2\pi{\rm i}\left[u^{2}_{\eta\vec{k}}\delta\left(\omega-{\rm E}_{\eta\vec{k}}-\mu_{\eta}\right)
+
v^{2}_{\eta\vec{k}}\delta\left(\omega+{\rm E}_{\eta\vec{k}}-\mu_{\eta}\right)\right]{\rm f}\left(\omega-\mu_{\eta}\right).
$$
therefore we have
\begin{eqnarray*}
V^{2}_{\eta}\sum_{\vec{k}}{\rm g}^{<}_{\eta\vec{k}\sigma}\left(\omega\right)
&=&
2\pi{\rm i}V^{2}_{\eta}\sum_{\vec{k}}\left[u_{\eta\vec{k}}^{2}\delta\left(\omega-{\rm
E}_{\eta\vec{k}}-\mu_{\eta}\right)+v^{2}_{\eta\vec{k}}\delta\left(\omega+{\rm E}_{\eta\vec{k}}-\mu_{\eta}\right)\right]\times
\\
&&{\rm f}\left(\omega-\mu_{\eta}\right),
\\
&=&
2\pi{\rm i}V^{2}_{\eta}\frac{1}{2}\sum_{\vec{k}}\delta\left(\left\arrowvert\omega-\mu_{\eta}\right\arrowvert-{\rm E}_{\eta\vec{k}}\right)
{\rm f}\left(\omega-\mu_{\eta}\right).
\end{eqnarray*}

Finally,

\fbox{\parbox{9cm}{
$$
{\rm \Sigma}^{<}\left(\omega\right)
\equiv
V^{2}_{\eta}\sum_{\vec{k}}{\rm g}^{<}_{\eta\vec{k}}\left(\omega\right)
=
2{\rm i}{\rm \Gamma}_{\eta}\zeta\left(\omega-\mu_{\eta},{\rm \Delta}_{\eta}\right){\rm f}\left(\omega-\mu_{\eta}\right).
$$
}}
\section{Appendix F}
\begin{eqnarray*}
{\rm f}^{({\rm r})}_{\eta\vec{k}\sigma}\left(t,t'\right)
&=&
-{\rm i}{\rm \Theta}\left(t-t'\right)\left\langle\left\{{\rm e}^{{\rm i}\mu_{\eta}t}a^{\dagger}_{\eta-\vec{k}-\sigma}\left(t\right), {\rm
e}^{{\rm i}\mu_{\eta}t'}a^{\dagger}_{\eta\vec{k}\sigma}\left(t'\right)\right\}\right\rangle
\\
&=&
-{\rm i}{\rm \Theta}\left(t+t'\right){\rm e}^{{\rm i}\mu_{\eta}\left(t+t'\right)}
\times\\
&&\left\langle\left\{u_{\eta\vec{k}}\gamma^{\dagger}_{\eta-\vec{k}-\sigma}\left(t\right)-\sigma
v_{\eta\vec{k}}\gamma_{\eta\vec{k}\sigma}\left(t\right), u_{\eta\vec{k}}\gamma^{\dagger}_{\eta\vec{k}\sigma}\left(t'\right) + \sigma
v_{\eta\vec{k}}\gamma_{\eta-\vec{k}-\sigma}\left(t'\right)\right\}\right\rangle
\nonumber\\
&=&
-{\rm i}{\rm \Theta}\left(t-t'\right){\rm e}^{2{\rm i}\mu_{\eta}t}{\rm e}^{-{\rm i}\mu_{\eta}\left(t-t'\right)}\sigma
v_{\eta\vec{k}}u_{\eta\vec{k}}\left[{\rm e}^{{\rm i}{\rm E}_{\eta\vec{k}}\left(t-t'\right)}-{\rm e}^{-{\rm i}{\rm E}_{\eta\vec{k}}\left(t-
t'\right)}\right]
\\
&=&
-{\rm i}{\rm \Theta}\left(t-t'\right)\frac{\sigma{\rm \Delta}_{\eta}}{2{\rm E}_{\eta\vec{k}}}\left[{{\rm e}^{{\rm i}\left({\rm
E}_{\eta\vec{k}}-\mu_{\eta}\right)\left(t-t'\right)}}-{{\rm e}^{-{\rm i}\left({\rm E}_{\eta\vec{k}}+\mu_{\eta}\right)\left(t-t'\right)}}
\right].
\end{eqnarray*}
Applying the Fourier transformations

$$
V^{2}_{\eta}\sum_{\vec{k}}{\rm f}^{({\rm r})}_{\eta\vec{k}\sigma}\left(t,t'\right)
=
\int^{\infty}_{-\infty}{\rm e}^{2{\rm i}\mu_{\eta}t}\sigma\frac{{\rm d}\omega}{2\pi}{\rm e}^{-{\rm i}\omega\left(t-t'\right)}{\rm
\Xi}^{({\rm r})}_{\eta\vec{k}}\left(\omega\right),
$$
with
\begin{eqnarray*}
V^{2}_{\eta}\sum_{\vec{k}}{\rm \Xi}^{({\rm r})}_{\eta\vec{k}}\left(\omega\right)
&=&
\frac{{\rm \Delta}_{\eta}}{2{\rm E}_{\eta\vec{k}}}\left[\frac{1}{\omega+{\rm E}_{\eta\vec{k}}-\mu_{\eta}+{\rm i}0^{+}}
-
\frac{1}{\omega- {\rm E}_{\eta\vec{k}}-\mu_{\eta}+{\rm i}0^{+}}\right]
\\
&=&
V^{2}_{\eta}\sum_{\vec{k}}\frac{{\rm \Delta}_{\eta}}{2{\rm E}_{\eta\vec{k}}}\left\lbrack\frac{1}{\omega+{\rm
E}_{\eta\vec{k}}-\mu_{\eta}+{\rm i}0^{+}}
-
\frac{1}{\omega-{\rm E}_{\eta\vec{k}}-\mu_{\eta}+{\rm i}0^{+}}\right\rbrack
\nonumber\\
&=&
V^{2}_{\eta}{\rm \Delta}_{\eta}\sum_{\vec{k}}\left\lbrack{\rm P}\frac{1}{\xi^{2}_{\eta\vec{k}}+{\rm
\Delta}^{2}_{\eta}-\left(\omega-\mu_{\eta}\right)^{2}}+
\frac{{\rm i}\pi}{2\left(\omega-\mu_{\eta}\right)}\delta\left(\left\arrowvert\omega-\mu_{\eta}\right\arrowvert-{\rm
E}_{\eta\vec{k}}\right)\right\rbrack
\nonumber\\
&=&
N_{\eta}\left(0\right)V^{2}_{\eta}{\rm \Delta}_{\eta}
\times\\
&&{\rm P}\int^{\infty}_{-\infty}{\rm d}\xi\frac{1}{\xi^{2}+{\rm \Delta}^{2}-\left(\omega-\mu\right)^{2}}
+
\frac{{\rm i}\pi V^{2}_{\eta}}
{2\left(\omega-\mu_{\eta}\right)}\sum_{\vec{k}}\delta\left(\left\arrowvert\omega-\mu_{\eta}\right\arrowvert-{\rm E}_{\eta\vec{k}}\right).
\end{eqnarray*}

Finally,

\fbox{\parbox{9cm}{
$$
{{\rm \Xi}}^{({\rm r})}_{\eta\vec{k}}\left(\omega\right)
=
{\rm \Gamma}_{\eta}\left\lbrack\zeta\left({\rm \Delta}_{\eta},\omega-\mu_{\eta}\right)
+
{\rm i}\frac{{\rm \Delta}_{\eta}}{\omega-\mu_{\eta}}\zeta\left(\omega-\mu_{\eta},{\rm \Delta}_{\eta}\right)\right\rbrack.
$$
}}

\section{Appendix G}
\begin{eqnarray*}
{\rm f}^{<}_{\eta\vec{k}\sigma}\left(t,t'\right)
&=&
{\rm i}\left\langle\left\{{\rm e}^{{\rm i}\mu_{\eta}t'}a^{\dagger}_{\eta\vec{k}\sigma}\left(t'\right), {\rm e}^{{\rm
i}\mu_{\eta}t}a^{\dagger}_{\eta-\vec{k}-\sigma}\left(t\right)\right\}\right\rangle
\\
&=&
{\rm i}{\rm e}^{{\rm i}\mu_{\eta}\left(t+t'\right)}\left\langle\left\lbrack
u_{\eta\vec{k}}\gamma^{\dagger}_{\eta\vec{k}\sigma}\left(t'\right)+\sigma v_{\eta
\vec{k}}\gamma_{\eta-\vec{k}-\sigma}\left(t'\right)\right\rbrack\right.
\times\\
&&\left.\left\lbrack u_{\eta\vec{k}}\gamma^{\dagger}_{\eta-\vec{k}-\sigma}\left(t\right) - \sigma v_{\eta
\vec{k}}\gamma_{\eta\vec{k}\sigma}\left(t\right)\right\rbrack\right\rangle
\\
&=&
{\rm i}{\rm e}^{2{\rm i}\mu_{\eta}t}{\rm e}^{-{\rm i}\mu_{\eta}\left(t-t'\right)}\left\lbrack\sigma u_{\eta\vec{k}}v_{\eta\vec{k}}{{\rm
e}^{{\rm i}\left({\rm E}_{\eta\vec{k}}\right)\left(t-t'\right)}}{\rm f}\left(-{\rm E}_{\eta\vec{k}}\right)\right.-
\\
&&\left.\sigma v_{\eta\vec{k}}u_{\eta\vec{k}}{{\rm e}^{-{\rm i}\left({\rm E}_{\eta\vec{k}}\right)\left(t-t'\right)}}{\rm f}\left({\rm
E}_{\eta\vec{k}}\right) \right\rbrack
\\
&=&
{\rm i}{\rm e}^{2{\rm i}\mu_{\eta}t}\sigma\frac{{\rm \Delta}_{\eta}}{2{\rm E}_{\eta\vec{k}}}\left\lbrack{{\rm e}^{{\rm i}\left({\rm
E}_{\eta\vec{k}}-\mu_{\eta}\right)\left(t-t'\right)}}{\rm f}\left(-{\rm E}_{\eta\vec{k}}\right)+{{\rm e}^{-{\rm i}\left({\rm
E}_{\eta\vec{k}}+\mu_{\eta}\right)\left(t-t'\right)}}{\rm f}\left({\rm E}_{\eta\vec{k}}\right) \right\rbrack.
\end{eqnarray*}
therefore we have
\begin{equation}
V^{2}_{\eta}\sum_{\vec{k}}{\rm f}^{<}_{\eta\vec{k}\sigma}\left(t,t'\right)
=
\int^{\infty}_{-\infty}{\rm e}^{2{\rm i}\mu_{\eta}t}\sigma\frac{{\rm d}\omega}{2\pi}{\rm e}^{-{\rm i}\omega\left(t-t'\right)}{{\rm
\Xi}}^{<}_{\eta\vec{k}}\left(\omega\right),
\label{appG1}
\end{equation}
with
\begin{eqnarray*}
{{\rm \Xi}}^{<}_{\eta\vec{k}}\left(\omega\right)
&=&
V^{2}_{\eta}\sum_{\vec{k}}\frac{\pi{\rm i}}{{\rm E}_{\eta\vec{k}}}\left\lbrack\delta\left(\omega+{\rm E}_{\eta\vec{k}}-\mu_{\eta}\right){\rm
f}\left(-{\rm E}_{\eta\vec{k}}\right)
-
\delta\left(\omega-{\rm E}_{\eta\vec{k}}-\mu_{\eta}\right){\rm f}\left({\rm E}_{\eta\vec{k}}\right)\right\rbrack
\\
&=&
V^{2}_{\eta}\sum_{\vec{k}}\frac{\pi{\rm i}}{{\rm E}_{\eta\vec{k}}}\left\lbrack\delta\left(\omega+{\rm E}_{\eta\vec{k}}-\mu_{\eta}\right)
-
\delta\left(\omega-{\rm E}_{\eta\vec{k}}-\mu_{\eta}\right)\right\rbrack{\rm f}\left(\omega-\mu_{\eta}\right)
\\
&=&
\frac{-2{\rm i}{\rm \Delta}_{\eta}}{\omega-\mu_{\eta}}\left\lbrack\frac{\pi V^{2}_{\eta}}
{2}\sum_{\eta}\delta\left(\left\arrowvert\omega-\mu_{\eta}\right\arrowvert-{\rm E}_{\eta\vec{k}}\right)\right\rbrack{\rm
f}\left(\omega-\mu_{\eta}\right).
\end{eqnarray*}

Finally,

\fbox{\parbox{8cm}{
$$
{{\rm \Xi}}^{<}_{\eta\vec{k}}\left(\omega\right)
=
-2{\rm i}{\rm \Gamma}_{\eta}\frac{{\rm \Delta}_{\eta}}{\omega-\mu_{\eta}}\zeta\left(\omega-\mu_{\eta},{\rm \Delta}_{\eta}\right)
{\rm f}\left(\omega-\mu_{\eta}\right).
$$
}}
%


\end{document}